\documentclass[fleqn,usenatbib]{mnras}

\usepackage[T1]{fontenc}

\DeclareRobustCommand{\VAN}[3]{#2}
\let\VANthebibliography\thebibliography
\def\thebibliography{\DeclareRobustCommand{\VAN}[3]{##3}\VANthebibliography}

\usepackage{graphicx}	% Including figure files
\usepackage{amsmath}	% Advanced maths commands
\usepackage{amssymb}	% Extra maths symbols

\usepackage{newtxtext,newtxmath}
\usepackage[dvipsnames]{xcolor}
\usepackage{enumitem}
\usepackage{multirow}

\usepackage{xcolor}

\makeatletter
\newcommand{\overleftrightsmallarrow}{\mathpalette{\overarrowsmall@\leftrightarrowfill@}}
\newcommand{\overrightsmallarrow}{\mathpalette{\overarrowsmall@\rightarrowfill@}}
\newcommand{\overleftsmallarrow}{\mathpalette{\overarrowsmall@\leftarrowfill@}}
\newcommand{\overarrowsmall@}[3]{%
  \vbox{%
    \ialign{%
      ##\crcr
      #1{\smaller@style{#2}}\crcr
      \noalign{\nointerlineskip}%
      $\m@th\hfil#2#3\hfil$\crcr
    }%
  }%
}
\def\smaller@style#1{%
  \ifx#1\displaystyle\scriptstyle\else
    \ifx#1\textstyle\scriptstyle\else
      \scriptscriptstyle
    \fi
  \fi
}
\makeatother

\newcommand{\svec}[1]{\boldsymbol{#1}}
\newcommand{\statevec}{\boldsymbol{u}}

\newcommand{\mean}[1]{\left\{\left\{#1\right\}\right\}}
\newcommand{\jump}[1]{\left[\left[#1\right]\right]}

\newcommand{\Secref}[1]{Section~\ref{#1}}
\newcommand{\Figref}[1]{Fig.~\ref{#1}}

\title[DG for FLASH]{A Discontinuous Galerkin Solver in the FLASH Multi-Physics Framework}

\author[J. Markert et al.]{
Johannes Markert,$^{1}$\thanks{Email: jmarkert@math.uni-koeln.de, Homepage: \url{www.jmark.de}}
Stefanie Walch,$^{2}$
Gregor Gassner$^{3}$
\\
$^{1}$ Department for Mathematics and Computer Science, University of Cologne, Weyertal 86-90, 50931, Cologne, Germany\\
$^{2}$ Universität zu Köln, Zülpicher Str. 77, I. Physikalisches Institut; Center for Data and Simulation Science 50937 Köln, Germany\\
$^{3}$ Department for Mathematics and Computer Science; Center for Data and Simulation Science, University of Cologne, Weyertal 86-90, 50931, Cologne, Germany\\
}

\date{Accepted XXX. Received YYY; in original form ZZZ}

\pubyear{2022}

\begin{document}
\label{firstpage}
\pagerange{\pageref{firstpage}--\pageref{lastpage}}
\maketitle

\begin{abstract}
In this paper, we present a discontinuous Galerkin solver based on previous
work by the authors for magneto-hydrodynamics in form of a new fluid
solver module integrated into the established and well-known multi-physics
simulation code FLASH. Our goal is to enable future research on the
capabilities and potential advantages of discontinuous Galerkin methods for
complex multi-physics simulations in astrophysical settings.  We give specific
details and adjustments of our implementation within the FLASH framework and
present extensive validations and test cases, specifically its interaction with
several other physics modules such as (self-)gravity and radiative transfer.
We conclude that the new DG solver module in FLASH is ready for use in
astrophysics simulations and thus ready for assessments and investigations.
\end{abstract}


%%%%%%%%%%%%%%%%% BODY OF PAPER %%%%%%%%%%%%%%%%%%

\section{Introduction}
In astrophysics, numerical simulations are a standard tool to study the
interplay of the complex physical processes which are shaping our universe.
As an example, understanding the star formation process requires numerical
simulations because the process is too slow to be followed in real time and the
star-forming gas is so dense that the young stellar objects are highly obscured
and cannot be directly observed. 
The star formation process involves compressible fluid dynamics with magnetic
fields and strong shocks, self-gravity of the gas, heating and cooling,
multi-component chemistry, and radiative transfer. 
It is therefore a challenging, multi-physics problem and requires accurate as
well as stable numerical methods and scalable high-performance computing
implementations.

There are many (open source) code frameworks in the astrophysics community that
include (a subset of) the aforementioned physical models with different
fidelity levels. A few examples are AREPO \citep{springel2010moving}, RAMSES
\citep{fromang2006high}, ENZO \citep{collins2010cosmological}, Dedalus \citep{burns2020dedalus}, ATHENA
\citep{stone2008athena}, ATHENA++ \citep{stone2020athena++}, ORION2 \citep{li2021orion2}, and the FLASH code \citep{fryxell2000flash}. 
Here, we list code frameworks that are based on discretisations with meshes, as
this strategy is also the focus of this paper. Most of these codes feature
adaptive mesh refinement (AMR), which is crucial to resolve the vastly
different spatial scales in the turbulent interstellar medium.  Furthermore,
most of these codes are based on Finite Volume (FV) type discretisations of the
fluid/plasma dynamics with numerical fluxes based on Riemann solvers at the
interface and, typically, second order reconstructions with slope limiting to
increase the accuracy by decreasing artificial dissipation.  In general, FV
methods are quite effective and robust and they are able to handle strong
shocks. Thus, they are currently the state-of-the-art in almost all grid-based
multi-physics astrophysical simulation codes. 

High-order discretisation schemes \citep[e.g.][]{balsara2017higher} promise a
higher fidelity and precision while being computationally efficient. This is
attractive since most astrophysical simulations are notoriously under-resolved
due to limited computing resources and/or parallel scalability of the employed
methods.  While there are many high-order methodologies available, this paper
focuses on discontinuous Galerkin (DG) schemes. DG schemes are a mixture of
high-order Finite Element methods with local polynomial basis functions and FV
methods in the sense that the ansatz space is discontinuous across grid cell
interfaces, which enables the use of Riemann solver based numerical fluxes.

The goal to efficiently make use of future exascale machines with their ever higher degree of parallel concurrency motivates the search for more efficient and more accurate techniques for computing hydrodynamics.  DG methods can be straightforwardly extended to arbitrarily high order of accuracy while requiring only minimal data from adjacent neighbours. Thus, the DG scheme has a very narrow data dependency with only a minimum of data exchange necessary, while the local compute kernels are computationally very dense. Furthermore, at least for subsonic turbulence, high-order DG offers significant computational benefits in computational efficiency for reaching a desired target accuracy, due to their spectral like very low dispersion and dissipation errors  \citep[e.g.][]{ainsworth2004dispersive,gassner2011comparison,dobrev2012high,he2020numerical}. Low artificial dissipation is important to reduce artificial damping and heating, in addition, low dispersion errors are equally important as it guarantees high accuracy for wave propagation and interaction. Last but not least, DG method intrinsically conserve angular momentum for approximation order 3 and higher \citep{despres2015angular}.

These beneficial properties of DG are the reason why applications in engineering are very successful
\citep[e.g.][]{Bassi2005,Zhang2010251,Moura2017,Manzanero2018b}, where mainly
weakly compressible turbulent flows in complex geometries are simulated. However, their
application in astrophysics for hypersonic flows is still in an early stage. DG methods with focus
on astrophysical fluid dynamics and related applications are for instance
presented in
\citet{schaal2015astrophysical,guillet2019high,Bauer2016a,KIDDER201784,10.1093/mnras/staa3682,mocz2014discontinuous}.

However, to the best of our knowledge, there is no DG implementation in a full
production multi-physics framework available up to now.  The scientific
contribution of this paper is to integrate a DG solver into an existing and
well-known multi-physics simulation code to enable research on the capabilities
and potential advantages of DG for complex multi-physics simulations in
astrophysical settings. Here we choose the FLASH code \citep{fryxell2000flash}.

To allow the robust simulation of highly supersonic, turbulent flows featuring
strong shocks, as is the case in most astrophysics models, proper shock
capturing for DG is mandatory. There are many shock capturing approaches available in the
DG literature, for instance based on slope limiters
\cite{Cockburn1998199,CockburnHou&Shu,Kuzmin2012} or (H)WENO limiters
\cite{Guo2015,Qiu2010,Zhu2009}, filtering \cite{Bohm2019aa}, and artificial
viscosity \cite{Persson2006,Ching2019}. Sub-cell based FV discretisations with a
first order, second order TVD or high order WENO reconstruction are introduced,
e.g., in \cite{Vilar2019,Sonntag2014,Dumbser2014,Zanotti2015,Dumbser2016}.

\cite{markert2021sub} introduced a shock capturing approach for
high-order DG which has shown to be robust under very strong astrophysical shock conditions (with large density and pressure contrasts), while retaining the beneficial properties of DG as much as possible and keeping physical quantities, like density and pressure, positive. As a side note, but important detail, the method of \cite{markert2021sub} is directly compatible with FLASH's block-based datastructure, and serves as a baseline approach for the current paper.

In the remainder of the paper we briefly discuss the governing equations used
for the DG implementation, as well as the specific details and adjustments of
the DG scheme by \cite{markert2021sub} that we have implemented in the FLASH
code. We then present extensive validations of the DG scheme and specifically
its interaction with several multi-physics modules. The idea is that we
increase the complexity of the test cases step by step, where the last test
case in \Secref{sec:fractal} is a more involved multi-physics astrophysics
application with multiple physics modules working together.

\section{Governing Equations}
\label{sec:gov-eqn}

We assume that the magnetised plasma in our astrophysical applications is
described reasonably by the equations of ideal magneto-hydrodynamics (iMHD)
\begin{equation}
\label{eq:ideal-MHD}
\partial_t \begin{pmatrix}
\rho \\ \rho\vec{v} \\ E \\ \vec{B}
\end{pmatrix} + \nabla \cdot \begin{pmatrix}
\rho\vec{v} \\ \rho\vec{v}\otimes\vec{v} + P\mathcal{I} - \vec{B}\otimes\vec{B} \\
(E + P)\vec{v} - (\vec{v}\cdot\vec{B})\vec{B} \\ \vec{B}\otimes\vec{v} - \vec{v}\otimes\vec{B}
\end{pmatrix} = 0,
\end{equation}
where $\rho$ is the density, $\vec{v} = (v_1,v_2,v_3)^T$ is the velocity, $E$
is the total energy, $\vec{B} = (B_1,B_2,B_3)^T$ is the magnetic field,
$\mathcal{I}$ represents the $3\times3$ identity matrix and the total pressure
$P$:
\begin{equation}
\label{eq:total-pressure}
P = p + \frac{1}{2}\vec{B}\cdot\vec{B}.
\end{equation}
The thermal pressure $p$ is given in equation \eqref{eq:thermal-pressure}.

A crucial property of magnetic fields is that their divergence is exactly zero
for all times. Numerically, it is challenging to exactly satisfy this condition
\citep[e.g.][]{brackbill1980effect,evans1988simulation,ryu1998divergence,balsara1999staggered,gardiner2005unsplit,dedner2002hyperbolic}.
In this work, we focus on an extension of the iMHD equations, that accounts for
the inexact divergence of the magnetic field by introducing a generalised
Lagrange multiplier (GLM), i.e., an additional equation that "carries" the
divergence error following the work of \citet{dedner2002hyperbolic}. However, we
focus on a specific variant of the GLM extension, that is entropy consistent
and consistent with the Lorentz force \citep{derigs2018ideal,Bohm2018}. In the
present implementation, we have further extended the MHD model to allow for
multiple components, necessary for the complex chemistry network interactions. 

\subsection{Ideal Generalised Lagrange Multiplier MHD}
\label{sec:ideal-GLM-MHD}

The form of the ideal generalised Lagrange multiplier magneto-hydrodynamics
(iGLM-MHD), extended for multiple chemical species and coupled to the gravity source
term reads as
\begin{equation}
\label{eq:balance-law}
    \partial_t \statevec + \partial_x \svec{F}(\statevec) + \partial_y \svec{G}(\statevec) + \partial_z \svec{H}(\statevec) 
    = -\svec{\Upsilon}^{\text{Powell}} - \svec{\Upsilon}^{\text{GLM}} + \svec{\Upsilon}^{\text{gravity}}.
\end{equation}
The vector of state variables (typeface in bold) consists of
\begin{equation}
\label{eq:conservative-state-vector}
\statevec = (\rho,\;\rho\,\vec{v},\;E,\;\vec{B},\;\Psi,\;\rho\,\svec{\sigma})^T \quad \in \mathbb{R}^{9+n}, 
\end{equation}
with the addition of the hyperbolic
divergence correction field $\Psi$ and the species vector $\svec{\sigma} =
(\sigma_1,\ldots,\sigma_n)^T$ describing mass fractions of the $n$ different chemical species. The thermal pressure, $p$, is computed from the total energy via
\begin{equation}
\label{eq:thermal-pressure}
    p = (\gamma-1)\Big(E - \frac{\rho}{2}\vec{v}\cdot\vec{v} - \frac{1}{2}\vec{B}\cdot\vec{B} - \frac{1}{2}\Psi^2\Big),
\end{equation}
where the heat capacity ratio, $\gamma$, is given by \eqref{eq:heat-capacity-ratio}.
The flux in x-direction reads as
\begin{equation}
\label{eq:xflux}
\svec{F} = \left(\begin{matrix}
    \rho v_1  \\[0.4em]
    \rho v_1 v_1 + p + \frac{1}{2}\vec{B}\cdot\vec{B} - B_1 B_1  \\[0.4em]
    \rho v_1 v_2 - B_1 B_2 \\[0.4em]
    \rho v_1 v_3 - B_1 B_3 \\[0.4em]
    v_1 \big(\frac{\rho}{2}\vec{v}\cdot\vec{v} + \frac{\gamma\,p}{\gamma-1} + \vec{B}\cdot\vec{B}\big) - B_1\,\vec{v}\cdot\vec{B} + c_h \Psi B_1\\[0.4em]
    c_h \Psi \\[0.4em]
    v_1 B_2 - v_2 B_1 \\[0.4em]
    v_1 B_3 - v_3 B_1 \\[0.4em]
    c_h B_1 \\[0.4em]
    \rho v_1 \svec{\sigma}
\end{matrix}\right) \in \mathbb{R}^{9+n},
\end{equation}
with the hyperbolic correction speed $c_h$ given by eq.~\eqref{eq:global-hyperbolic-correction-speed}. The fluxes in $y$- and
z-direction, respectively $G$ and $H$, are listed in \citet{derigs2018ideal}
Appendix F, yet without the multi-species extension.

What follows is a detailed discussion of the different parts of the governing
equations: quasi-multi-fluids, maximum wave speeds, Powell source terms
$\svec{\Upsilon}^{\text{Powell}}$ and the hyperbolic divergence cleaning technique
with its associated source term $\svec{\Upsilon}^{\text{GLM}}$.

\subsubsection{Quasi-multi-fluid Model}
\label{sec:quasi-multifluid}
The ability to track the exact composition of a fluid or gas is of central
importance in astrophysical simulations as they include detailed chemical
reaction chains (chemical networks) to treat heating, cooling, as well as the
formation and destruction of chemical compounds in order to mimic the behaviour
of the interstellar medium \citep[ISM; ][]{walch2015silcc,gatto2015modelling,glover2014molecular}.
The individual species are traced by mass fractions $\sigma_s \in [0,1]$, which all move
with the same velocity $\vec{v}$ as the total density $\rho$:
\begin{equation}
\label{eq:quasi-multifluid}
    \partial_t (\rho\,\sigma_s) + \nabla\cdot(\rho\,\vec{v}\,\sigma_s) = 0, \quad s = 1,\ldots,n.
\end{equation}
The sum of all $n$ mass fractions maintains the total density at all times,
i.e. $\sum_s^n\sigma_s = 1$.

We generalise our scheme for a multi-species fluid with a variable heat
capacity ratio $\gamma$ by adopting a weighted mean over all species
\citep{murawski2002analytical}:
\begin{equation}
\label{eq:heat-capacity-ratio}
\gamma = \frac{\sum_s^n c^{\text{pres.}}_s\,\sigma_s}{\sum_s^n c^{\text{vol.}}_s\,\sigma_s}
\end{equation}
with the heat capacities for constant pressure $c^{\text{pres.}}_s$ and
constant volume $c^{\text{vol.}}_s$ of each individual species. The standard
FLASH framework readily offers inbuilt multi-species support with modules taking care of
the correct equation-of-state calculations \eqref{eq:heat-capacity-ratio}. 
It considerably minimises the implementation effort on our side by only focusing on the
proper species advection part.

In addition to the multi-fluid approach, we also support mass tracer fields (also
called mass scalars) which are advected similar to \eqref{eq:quasi-multifluid}.
The implementation of the tracer fields allows the use of any number of such
fields which makes it a flexible tool for tracing different mass quantities
according to individual requirements. For example, a mass tracer field could be
used to follow the distribution of metals in the interstellar gas with
virtually no additional costs. Note, that we do not include mass tracer fields
explicitly in \eqref{eq:conservative-state-vector} since they do not involve
any special numerical discretisation than already presented here.

\subsubsection{Maximum Wave Speed}
\label{sec:signal-speeds}
A vital step in our numerical treatment of iGLM-MHD is to compute the maximum
eigenvalue of the system. It encodes the maximum wave speed involved in the
solution and helps to find a good estimation for an acceptable timestep in case
of explicit time integration. A thorough investigation of the full eigenvalue
system is presented in \citet{derigs2018ideal}. In this work, we focus on the fastest
signal called the fast magneto-sonic wave speed. In direction $d = 1,2,3$ it is
calculated as
\begin{equation}
\label{eq:signal-speed}
    c_d = \frac{\sqrt{c_{\text{sound}}^2 + c_{\text{Alfvén}}^2 + \sqrt{(c_{\text{sound}}^2 + c_{\text{Alfvén}}^2)^2 - (2\,c_{\text{sound}}B_d/\sqrt{\rho})^2}}}{\sqrt{2}},
\end{equation}
where the sonic wave speed and Alfvén wave speed read
\begin{equation}
\label{eq:wave-speeds}
    c_{\text{sound}} = \sqrt{\gamma\,\frac{p}{\rho}} \quad \text{and} \quad
    c_{\text{Alfvén}} = \frac{|\vec{B}|}{\sqrt{\rho}}.
\end{equation}
The upper bound of all involved wave speeds in direction $d = 1,2,3$ is
then given by
\begin{equation}
\label{eq:mhd-max-wave-speed}
\lambda^{\text{max}}_d = \big|v_d\big| + c_d.
\end{equation}

\subsubsection{Powell Source Terms}
Following the theoretical groundwork by \citet{godunov1972symmetric},
\citet{powell1999solution} pointed out that the system \eqref{eq:ideal-MHD} is
not Galilean invariant and does not formally conserve entropy. They proposed to
add a specific source term proportional to $\nabla \cdot B$ in order to
symmetrize the system. The so-called Powell terms can be obtained from deriving
the local form of the system \eqref{eq:ideal-MHD} based on integral
conservation laws \citep{powell1999solution} or from requiring entropy stability
\citep{godunov1972symmetric,winters2016entropy,chandrashekar2016entropy,derigs2018ideal}.

We write the Powell source terms in eq.~\eqref{eq:balance-law} as
\begin{equation}
\label{eq:powell-source-term}
\svec{\Upsilon}^{\text{Powell}} = \big(\partial_x B_1 + \partial_y B_2 + \partial_z B_3\big)\;\svec{\Phi}^{\text{Powell}} \quad\text{with}
\end{equation}
\begin{align}
\label{eq:powell-source-vectors}
\svec{\Phi}^{\text{Powell}} &= \big(0,\;\vec{B},\; \vec{v}\cdot\vec{B},\; \vec{v},\; 0,\; \svec{0}\big)^T \quad \in \mathbb{R}^{9+n}.
\end{align}

The Powell method can be easily adapted to Eulerian grid codes
\citep{balsara1999staggered,toth2000b}. It has been successfully implemented
and tested in astrophysical MHD codes equipped with a FV scheme and
AMR \citep{derigs2016novel}. For DG it is for instance adopted by
\citet{warburton1999discontinuous} and \citet{Bohm2018} for viscous and resistive MHD flows.

The Powell method has two limitations. Firstly, it does not fully eliminate the
divergence error and can result in a local accumulation of numerical divergence
in regions of stagnant flows. Secondly, the source term is not strictly
conservative anymore since it will locally inject conserved quantities in the
presence of numerical divergence errors. We investigate these issues in the
numerical results in \Secref{sec:current-sheet} and argue that the combination with
hyperbolic divergence cleaning, discussed in the next paragraph, remedies these
issues and leads to acceptable results.

\subsubsection{Hyperbolic Divergence Cleaning}
\label{sec:hyperbolic-div-cleaning}
\citet{munz2007linearized} coupled the divergence constraint for the electric
field with the induction equation by introducing a generalised Lagrangian
multiplier (GLM), $\Psi$, as an additional field. As in
\citet{dedner2002hyperbolic}, we apply this technique to iMHD in order to account
for the divergence-free condition $\nabla \cdot B = 0$ by adding the GLM $\Psi$
as another conservative state variable, which we call hyperbolic divergence
correction field. This new field couples to the divergence of the magnetic
field through a modified induction equation \citep{derigs2018ideal}:
\begin{equation*}
\partial_t\vec{B} + \nabla\cdot(\vec{B}\otimes\vec{v} - \vec{v}\otimes\vec{B}) + \nabla\cdot\Psi = 0.
\end{equation*}
The field $\Psi$ evolves via the dynamical equation
\begin{equation*}
\partial_t\Psi + c_h\nabla\cdot\vec{B} + \vec{v}\,\nabla\Psi + c_p\Psi = 0,
\end{equation*}
resulting in a coupled GLM-iMHD system, which makes the fluctuations of $\Psi$
propagate away from their sources at speed $c_h > 0$ while damping them with
the damping speed $c_p > 0$ at time-scales $\propto c_h^{-1}$.

We write the GLM source terms in eq.~\eqref{eq:balance-law} as
\begin{equation}
\label{eq:GLM-source-term}
\svec{\Upsilon}^{\text{GLM}} = \big(v_1\;\partial_x \Psi + v_2\;\partial_y \Psi + v_3\;\partial_z \Psi\big)\;\svec{\Phi}^{\text{GLM}} + c_p\,\svec{\Phi}^{\text{damp}},
\end{equation}
with GLM vectors
\begin{align}
\label{eq:GLM-source-vectors}
\svec{\Phi}^{\text{GLM}}    &= \big(0,\;\vec{0},\; \Psi,\; \vec{0},\; 1,\; \svec{0}\big)^T \in \mathbb{R}^{9+n} \quad\text{and} \\
\label{eq:GLM-source-damp}
\svec{\Phi}^{\text{damp}}   &= \big(0,\;\vec{0},\; 0,\; \vec{0},\; \Psi,\; \svec{0}\big)^T \in \mathbb{R}^{9+n}.
\end{align}

For efficient divergence propagation without compromising stability we set
$c_h$ to be the maximum magneto-sonic wave speed (see eq.~\eqref{eq:signal-speed}) present
in the entire physical domain $\Omega$:
\begin{equation}
\label{eq:global-hyperbolic-correction-speed}
    c_h = \max_{\Omega}\max_{d=1}^{3} c_d. %\;-\; \max_{\Omega}\max_{d=1}^{3} |v_d|.
\end{equation}
According to \citet{dedner2002hyperbolic} a good choice for
the damping speed, $c_p$, is given by
\begin{equation*}
    c_p = \frac{c_h}{0.18}.
\end{equation*}

The GLM method is straightforward to implement in existing schemes and it has
also been adopted by a number of MHD codes
\citep{gaburov2011astrophysical,mignone2012conservative,dumbser2016simple,Bohm2018,rueda2021entropy}.

\subsubsection{Coupling to Gravity}
\label{sec:coupling-to-gravity}
The inclusion of gravity in the governing equations introduces a force on the
right-hand side of the momentum equations of the form
\begin{equation*}
\partial_t(\rho\vec{v}) + \rho\nabla\phi = 0,
\end{equation*}
where the gravitational potential $\phi$ satisfies the Poisson equation
\begin{equation*}
\nabla^2\phi = 4\pi\,G\,\rho,
\end{equation*}
with gravitational constant $G$. The source term in eq.~\eqref{eq:balance-law} then
reads
\begin{equation}
\label{eq:gravity-source-term}
\svec{\Upsilon}^{\text{gravity}} = \big(0,\; \rho\,\vec{g},\; \rho\,\vec{v}\cdot\vec{g},\; \vec{0},\; 0,\; \svec{0}\big)^T \in \mathbb{R}^{9+n},
\end{equation}
with the gravitational acceleration $\vec{g} = (g_1,g_2,g_3)^T$. The gravity
accelerations are computed with any of the gravity solvers available in the
FLASH framework \citep[e.g.][]{wunsch2018tree}. We expand on available gravity solvers in
\Secref{sec:coupling-gravity}.

\section{Numerical Discretisation}
\label{sec:numerical-discretization}
In this section, we give a brief description of our numerical discretisation of
the system given in eq.~\eqref{eq:balance-law}. We employ a newly developed DG
variant based on blocks of mean values \citep{markert2021sub}. This enables us
to construct a shock capturing approach based on the idea that we use
high-order DG whenever possible, whereas we blend the scheme with a
second-order slope limiting FV scheme when there are for instance strong
shocks. 

The standard solver in FLASH is based on an AMR enabled second-order FV method,
where the fluid variables are stored in the form of mean values.  FLASH
organises by default the FV mean values in blocks of specific sizes, e.g., of
size $8\times 8\times 8$. It is important to note, that all other physics
modules of FLASH assume that the data is organised in form of such blocks with
mean values and hence that the interaction between the additional physics
modules and the FLASH fluid solvers is based on mean values. In contrast,
typically, in DG the fluid variables are stored in form of local polynomials
with either modal coefficients \citep{Karniadakis:2005fj} or nodal values
\citep{kopriva2009implementing}. Depending on the polynomial degree $N-1$ of the
local ansatz, $N^3$ unknowns per fluid variable form a data package. Thus, the
size of the blocks (or in DG jargon size of the "elements"), varies with the
choice of the polynomial degree.

This difference in the representation of the solutions and the interpretation
of the solution coefficients forms a tough challenge that needs to be overcome.
Roughly, we have to major choices: (i) We can implement the DG scheme in its
typical/natural form based on the polynomial representations. This means,
however, that we than need to adjust the whole FLASH code to cater to the
piece-wise polynomial data. Thus, changes in the way the grid is managed (not
blocks of constant size, but smaller elements) and changes in the way all other
physics modules interact with the DG solver have to be changed and implemented.
(ii) We adapt the DG methodology and specifically design a variant that
directly operates with the block based mean value data structure of FLASH. The
upside is of course that in this case DG can with reasonable implementation
effort directly access all the functionality that the FLASH code offers. The
downside is that this is not the "natural" way one would implement the DG
methodology and that there are thus some disadvantages, such as additional
transforms between ansatz spaces, that one needs to accept.

In this paper, we have decided in favour of option (ii), where we adjust the DG
methodology such that we can directly use the rich multi-physics framework that
FLASH offers. In previous work, we have presented a robust and accurate DG
variant that uses blocks of mean values as input \citep{markert2021sub}, which
serves as a starting reference for the FLASH implementation.

Besides option (ii) being presumably the only feasible approach regarding the
necessary amount of implementation, it further has beneficial effects such as
being able to directly use the post-processing tool chains for data organised
in blocks of mean values, which have been established in many years of research
work.

We first introduce the notations for FV and DG separately, and then we combine
both schemes via convex blending. We also highlight our solution to make the
two schemes compatible within the FLASH framework while maintaining the key
numerical properties such as conservation and accuracy. We conclude this
section with a short discussion about the employed time integration method and
how we enforce positivity of density and pressure.

\subsection{Finite Volume Scheme}
\label{sec:finite-volume-scheme}
We assume a Cartesian grid and subdivide the physical domain $\Omega$ into $Q$
blocks, each consisting of $N \times N \times N$ regular sub-cells of size
\begin{equation*}
\left(\frac{\Delta x_q}{N}, \frac{\Delta y_q}{N}, \frac{\Delta z_q}{N}\right)^T =: \frac{\Delta \vec{x}_q}{N}.
\end{equation*}
Each sub-cell of block $q$ represents the cell-averaged fluid state
$\svec{u}(t,\vec{x})$ at sub-cell centres $\vec{x} = \vec{\mu}_{q,ijk}$ and time
$t$:
\begin{equation}
\label{eq:mean-values}
\overline{\svec{u}}_{q,ijk}(t) = \frac{N^3}{\Delta x_q \, \Delta y_q \, \Delta z_q} \iiint_{\vec{\mu}_{q,ijk}-\frac{\Delta \vec{x}_q}{2\,N}}^{\vec{\mu}_{q,ijk}+\frac{\Delta \vec{x}_q}{2\,N}} \svec{u}(t,\vec{x})\,dxdydz,
\end{equation}
with $i,j,k = 1,\ldots,N$. For the sake of brevity, we do not declare the range
of indices $i,j,k$ in each equation from here on.  Furthermore, we write a bar
over each quantity that represents a cell-averaged or face-averaged mean value.

Following the standard procedure \citep{Toro:1999yq}, we construct within each
block $q$ a three-dimensional FV scheme. In semi-discrete form it reads
\begin{align}
\label{eq:3D-fv-scheme}
\partial_t \overline{\svec{u}}_{q,ijk}^{\text{FV}} =
    &-\frac{N}{\Delta x_q}\big(\overline{\svec{F}}^*_{q,i+\frac{1}{2}jk} - \overline{\svec{F}}^*_{q,i-\frac{1}{2}jk}\big)  \nonumber\\
    &-\frac{N}{\Delta y_q}\big(\overline{\svec{G}}^*_{q,ij+\frac{1}{2}k} - \overline{\svec{G}}^*_{q,ij-\frac{1}{2}k}\big) \nonumber \\
    &-\frac{N}{\Delta z_q}\big(\overline{\svec{H}}^*_{q,ijk+\frac{1}{2}} - \overline{\svec{H}}^*_{q,ijk-\frac{1}{2}}\big) \nonumber \\
    &- \overline{\svec{\Upsilon}}^{\text{Powell}}_{q,ijk} 
     - \overline{\svec{\Upsilon}}^{\text{GLM}}_{q,ijk}.
     %+ \overline{\svec{\Upsilon}}^{\text{gravity}}_{q,ijk}.
\end{align}
The consistent numerical flux $\svec{F}^*$ has the general form
\begin{align}
\label{eq:numerical-flux}
&\overline{\svec{F}}^*_{q,i+\frac{1}{2}jk} = \svec{F}^*\big(\svec{u}^+_{q,i+\frac{1}{2}jk},\svec{u}^-_{q,i+\frac{1}{2}jk}\big) \\
&\:= \svec{F}^{\text{KEPEC}}\big(\svec{u}^+_{q,i+\frac{1}{2}jk},\svec{u}^-_{q,i+\frac{1}{2}jk}\big) 
        - \text{\textbf{Stab}}\big(\svec{u}^+_{q,i+\frac{1}{2}jk},\svec{u}^-_{q,i+\frac{1}{2}jk}\big)\nonumber, 
\end{align}
where $\svec{F}^{\text{KEPEC}}$ is a kinetic energy and entropy conserving
(KEPEC) two-point flux derived in \citet{derigs2018ideal}. The interface values
$\svec{u}^+$ and $\svec{u}^-$ are retrieved from a total variation diminishing
(TVD) reconstruction method. A thorough discussion about time-proven
reconstruction methods can be found in \citet{Toro:1999yq}.

In order to introduce dissipation the numerical flux is subtracted with a
stabilisation term \textbf{Stab}. In this work we use the simple and very
robust Rusanov-type dissipation:
\begin{equation}
\label{eq:rusanov-flux}
\overline{\svec{F}}^{*}_{q,i+\frac{1}{2}jk} 
    = \overline{\svec{F}}^{\text{KEPEC}}_{q,i+\frac{1}{2}jk}
    - \frac{1}{2}\lambda^{\text{max}}_{q,i+\frac{1}{2}jk}\Big(\svec{u}^+_{q,i+\frac{1}{2}jk}-\svec{u}^-_{q,i+\frac{1}{2}jk}\Big),
\end{equation}
with
\begin{equation*}
\lambda^{\text{max}}_{q,i+\frac{1}{2}jk} =
\max\Big\{\lambda^{\text{max}}_1(\svec{u}^+_{q,i+\frac{1}{2}jk}),\;\lambda^{\text{max}}_1(\svec{u}^-_{q,i+\frac{1}{2}jk})\Big\}
\end{equation*}
being the maximum wave speed \eqref{eq:mhd-max-wave-speed} at the interface.
Since the KEPEC flux in \citet{derigs2018ideal} is not specified for multiple
species we resort to the following advection scheme
\begin{equation}
\label{eq:rusanov-flux-species}
\overline{\svec{F}}^{*,\text{species}}_{q,i+\frac{1}{2}jk} 
    = \mean{\rho v_1 \svec{\sigma}}_{q,i+\frac{1}{2}jk}
    - \frac{1}{2}\lambda^{\text{max}}_{q,i+\frac{1}{2}jk}\jump{\rho \svec{\sigma}}_{q,i+\frac{1}{2}jk}.
\end{equation}
To allow a succinct notation we introduce the mean and jump operators, respectively,
\begin{align}
\mean{\cdot}_{q,i+\frac{1}{2}jk} &= \frac{1}{2}\Big((\cdot)^+_{q,i+\frac{1}{2}jk} + (\cdot)^-_{q,i+\frac{1}{2}jk}\Big), \quad \text{and} \\
\jump{\cdot}_{q,i+\frac{1}{2}jk} &= (\cdot)^-_{q,i+\frac{1}{2}jk} - (\cdot)^+_{q,i+\frac{1}{2}jk}.
\end{align}
The fluxes $\svec{G}^*$ and $\svec{H}^*$ are defined analogously.

The gradients in the Powell term \eqref{eq:powell-source-term} are approximated
with central finite differencing over sub-cell $\overline{\svec{u}}_{q,ijk}$.
In x-direction we write
\begin{equation}
\label{eq:powell-fv}
\overline{\svec{\Upsilon}}^{\text{Powell},x}_{q,ijk} 
    = \frac{N}{\Delta x_q}\Big(\mean{B_1}_{q,i+\frac{1}{2}jk} - \mean{B_1}_{q,i-\frac{1}{2}jk}\Big)\;\overline{\svec{\Phi}}^{\text{Powell}}_{q,ijk},
\end{equation}
where the source vector is calculated by inserting the mean values into
expression \eqref{eq:powell-source-vectors}:
$\overline{\svec{\Phi}}^{\text{Powell}}_{q,ijk} :=
\svec{\Phi}^{\text{Powell}}\big(\overline{\svec{u}}_{q,ijk}\big)$.
The complete flux is then the sum along all directions
\begin{equation*}
\overline{\svec{\Upsilon}}^{\text{Powell}}_{q,ijk} = \overline{\svec{\Upsilon}}^{\text{Powell},x}_{q,ijk} +
\overline{\svec{\Upsilon}}^{\text{Powell},y}_{q,ijk} + \overline{\svec{\Upsilon}}^{\text{Powell},z}_{q,ijk}.
\end{equation*}
The discretisation of the GLM term \eqref{eq:powell-source-term} is done analogously and reads
\begin{equation}
\label{eq:GLM-fv}
\overline{\svec{\Upsilon}}^{\text{GLM},x}_{q,ijk} 
    = v_1\;\frac{N}{\Delta x_q}\Big(\mean{\psi}_{q,i+\frac{1}{2}jk} - \mean{\psi}_{q,i-\frac{1}{2}jk}\Big)\;\overline{\svec{\Phi}}^{\text{GLM}}_{q,ijk}.
\end{equation}
The complete flux is the sum along all directions and the damping term:
\begin{equation*}
\overline{\svec{\Upsilon}}^{\text{GLM}}_{q,ijk} = \overline{\svec{\Upsilon}}^{\text{GLM},x}_{q,ijk} +
\overline{\svec{\Upsilon}}^{\text{GLM},y}_{q,ijk} + \overline{\svec{\Upsilon}}^{\text{GLM},z}_{q,ijk} + c_p \overline{\svec{\Phi}}^{\text{damp}}_{q,ijk}.
\end{equation*}

\subsection{Discontinuous Galerkin Spectral Element Method}
\label{sec:dgsem}
Similar to the previous section we assume a Cartesian grid and subdivide the
physical domain $\Omega$ into $Q$ elements of size
\begin{equation*}
\left(\Delta x_q, \Delta y_q, \Delta z_q\right)^T =: \Delta \vec{x}_q.
\end{equation*}
Inside each element $q$ we make a polynomial tensor product ansatz of degree
$N-1$ and approximate the exact solution $\svec{u}(t,\vec{x})$ by
\begin{equation}
\label{eq:polynomial-ansatz}
\svec{u}\big(t,\vec{x}_q(\vec{\chi})\big) \approx \sum^N_{ijk=1} \svec{u}\big(t,\vec{x}_q(\vec{\xi}_{ijk})\big) \; \ell_i(\chi_1)\ell_j(\chi_2)\ell_k(\chi_3),
\end{equation}
where $\vec{x}_q(\vec{\chi}) = \vec{\mu}_q +  \vec{\chi}\,\Delta\vec{x}_q \in
\Omega$ is a mapping from reference space $\vec{\chi} \in
\big[-\frac{1}{2},\frac{1}{2}\big]^3$ to the physical domain anchored around
the element's midpoint $\vec{\mu}_q$. The \textsc{Lagrange} polynomials
$\ell_i,\ell_j,\ell_k$ are pinned to the \textsc{Legendre-Gauss} quadrature
nodes 
\begin{equation}
\label{eq:quadrature-nodes}
    \vec{\xi}_{ijk} = (\xi_i,\xi_j,\xi_k)^T \in \Big[-\frac{1}{2},\frac{1}{2}\Big]^3, \quad i = 1,\ldots,N.
\end{equation}
The associated quadrature weights, $\omega_i \in \big(0,1\big)$, fulfil
$\sum^N_i \omega_i = 1$.

Following the standard procedure in \citet{kopriva2009implementing}, we get the
semi-discrete weak-form 
\begin{align}
\label{eq:3D-dg-scheme}
&\partial_t \widetilde{\svec{u}}_{q,ijk}^{\text{DG}} = \\
    &\quad-\frac{1}{\omega_i\,\Delta x_q}\Big(\mathcal{B}^+_{i}\,\widetilde{\svec{F}}^{*}_{q+\frac{1}{2},jk} - \mathcal{B}^-_{i}\,\widetilde{\svec{F}}^{*}_{q-\frac{1}{2},jk} - \sum^N_{l=1} \omega_l\,\mathcal{D}_{li}\,\widetilde{\svec{F}}_{q,ljk}\Big) \nonumber \\
    &\quad-\frac{1}{\omega_j\,\Delta y_q}\Big(\mathcal{B}^+_{j}\,\widetilde{\svec{G}}^{*}_{q+\frac{1}{2},ik} - \mathcal{B}^-_{j}\,\widetilde{\svec{G}}^{*}_{q-\frac{1}{2},ik} - \sum^N_{l=1} \omega_l\,\mathcal{D}_{lj}\,\widetilde{\svec{G}}_{q,ilk}\Big) \nonumber \\
    &\quad-\frac{1}{\omega_k\,\Delta z_q}\Big(\mathcal{B}^+_{k}\,\widetilde{\svec{H}}^{*}_{q+\frac{1}{2},ij} - \mathcal{B}^-_{k}\,\widetilde{\svec{H}}^{*}_{q-\frac{1}{2},ij} - \sum^N_{l=1} \omega_l\,\mathcal{D}_{lk}\,\widetilde{\svec{H}}_{q,ijl}\Big) \nonumber \\
    &\quad-\widetilde{\svec{\Upsilon}}^{\text{Powell}}_{q,ijk} - \widetilde{\svec{\Upsilon}}^{\text{GLM}}_{q,ijk},
\end{align}
where the differentiation operator is constructed by
\begin{equation}
\label{eq:diffmatrix}
    \mathcal{D}_{ij} = \partial_{\xi} \ell_j(\xi)\,\Big|_{\xi_i}, \quad i,j = 1,\ldots,N,
\end{equation}
and the boundary interpolation operators read
\begin{equation}
\label{eq:surfvector}
    \mathcal{B}^{\pm}_{i} = \ell_i\left(\pm\frac{1}{2}\right), \quad i = 1,\ldots,N.
\end{equation}
We adopt the notation in \citet{markert2021sub} and annotate quantities in nodal
space with the tilde sign. Note that the notation for indexing the element
interfaces, i.e. $q\pm\frac{1}{2}$, is to be understood in an abstract sense.
Thus, $\widetilde{\svec{F}}^{*}_{q\pm\frac{1}{2}}$,
$\widetilde{\svec{G}}^{*}_{q\pm\frac{1}{2}}$ and
$\widetilde{\svec{H}}^{*}_{q\pm\frac{1}{2}}$ point to the faces in x-, y- and
z-direction, respectively.

The nodal volume flux
\begin{equation}
\label{eq:volume-flux}
    \widetilde{\svec{F}}_{q,ijk} = \svec{F}\big(\widetilde{\svec{u}}_{q,ijk}\big) \quad \text{with} \quad
\widetilde{\svec{u}}_{q,ijk} := \svec{u}\big(t,\vec{x}_q(\vec{\xi}_{ijk})\big)
\end{equation}
is computed from the polynomial coefficients given in eq.~\eqref{eq:polynomial-ansatz}. The surface flux
\begin{equation}
\label{eq:surface-flux}
\widetilde{\svec{F}}^{*}_{q\pm\frac{1}{2},jk} = \svec{F}^*\big(\widetilde{\svec{u}}^{+}_{q\pm\frac{1}{2},jk},\widetilde{\svec{u}}^{-}_{q\pm\frac{1}{2},jk}\big)
\end{equation}
is calculated analogously to the FV scheme with the Rusanov flux
\eqref{eq:rusanov-flux} and the interpolated element boundary values
\begin{equation}
\label{eq:dg-boundary-interpolation}
    \widetilde{\svec{u}}^{+}_{q+\frac{1}{2},jk} = \sum^N_{i=1} \mathcal{B}^{+}_{i} \; \widetilde{\svec{u}}_{q,ijk} \quad\text{and}\quad
    \widetilde{\svec{u}}^{-}_{q+\frac{1}{2},jk} = \sum^N_{i=1} \mathcal{B}^{-}_{i} \; \widetilde{\svec{u}}_{q+1,ijk}
\end{equation}
at the common interface $q+\frac{1}{2}$ between neighbouring elements $q$ and
$q+1$. The computations in y- and z-direction are done analogously.

We split the discretisation of the Powell term \eqref{eq:powell-source-term}
into a surface and volume part. In x-direction we get
\begin{align}
\label{eq:powell-dg-surf-vol}
\widetilde{\svec{\Upsilon}}^{\text{Pow,x,surf}}_{q,ijk}
    &= \frac{1}{\omega_i\,\Delta x_q}\Big(\mathcal{B}^+_{i}\,\mean{\widetilde{B}_1}^+_{q,jk} - \mathcal{B}^-_{q,i}\,\mean{\widetilde{B}_1}^-_{q,jk} 
    \Big)\,\widetilde{\svec{\Phi}}^{\text{Powell}}_{q,ijk},\nonumber\\
\widetilde{\svec{\Upsilon}}^{\text{Pow,x,vol}}_{q,ijk} &= -\frac{1}{\omega_i\,\Delta x_q}\Big( 
    \sum^N_{l=1} \omega_l\,\mathcal{D}_{li}\,(\widetilde{B}_1)_{q,ljk} \Big)\,\widetilde{\svec{\Phi}}^{\text{Powell}}_{q,ijk}.
\end{align}
The source vector is calculated by inserting the nodal states
$\widetilde{\svec{u}}_{q,ijk}$ into expression \eqref{eq:powell-source-vectors}:
$\widetilde{\svec{\Phi}}^{\text{Powell}}_{q,ijk} :=
\svec{\Phi}^{\text{Powell}}\big(\widetilde{\svec{u}}_{q,ijk}\big)$. Furthermore,
we calculate element interface averages
\begin{equation}
\label{eq:block-interface-average}
    \mean{\cdot}^{\pm}_{q,jk} = \frac{1}{2}\Big((\cdot)^+_{q\pm\frac{1}{2},jk} + (\cdot)^-_{q\pm\frac{1}{2},jk}\Big)
\end{equation}
from the interpolated interface values \eqref{eq:dg-boundary-interpolation}.
The full Powell term is the sum of the volume and surface contributions from
all directions:
\begin{equation}
\label{eq:powell-dg}
\widetilde{\svec{\Upsilon}}^{\text{Powell}}_{q,ijk} = \widetilde{\svec{\Upsilon}}^{\text{Pow,x,surf}}_{q,ijk} + \widetilde{\svec{\Upsilon}}^{\text{Pow,x,vol}}_{q,ijk}
+ \widetilde{\svec{\Upsilon}}^{\text{Pow,y,surf}}_{q,ijk} + \ldots
\end{equation}

Following the same approach as for the Powell term, the GLM term \eqref{eq:GLM-source-term} then reads
\begin{align}
\label{eq:glm-dg-surf-vol}
\widetilde{\svec{\Upsilon}}^{\text{GLM,x,surf}}_{q,ijk}
    &= \frac{(\widetilde{v}_1)_{q,ijk}}{\omega_i\,\Delta x_q}\Big(\mathcal{B}^+_{i}\,\mean{\widetilde{\Psi}}^+_{q,jk} - \mathcal{B}^-_{i}\,\mean{\widetilde{\Psi}}^-_{q,jk} 
    \Big)\,\widetilde{\svec{\Phi}}^{\text{GLM}}_{q,ijk}, \nonumber \\
\widetilde{\svec{\Upsilon}}^{\text{GLM,x,vol}}_{q,ijk} &= -\frac{(\widetilde{v}_1)_{q,ijk}}{\omega_i\,\Delta x_q}\Big( 
    \sum^N_{l=1} \omega_l\,\mathcal{D}_{li}\,\widetilde{\Psi}_{q,ljk} \Big)\,\widetilde{\svec{\Phi}}^{\text{GLM}}_{q,ijk}.
\end{align}
Again, the full GLM source term is the sum of all contributions from all directions plus
the damping term
\begin{equation}
\label{eq:glm-dg}
\widetilde{\svec{\Upsilon}}^{\text{GLM}}_{q,ijk} = \widetilde{\svec{\Upsilon}}^{\text{GLM,x,surf}}_{q,ijk} + \widetilde{\svec{\Upsilon}}^{\text{GLM,x,vol}}_{q,ijk}
+ \widetilde{\svec{\Upsilon}}^{\text{GLM,y,surf}}_{q,ijk} + \ldots + c_p\,\widetilde{\svec{\Phi}}^{\text{damp}}_{q,ijk}.
\end{equation}

\subsection{Blending Scheme}
\label{sec:blending-scheme}
\begin{figure}
\centering
\includegraphics[width=1.0\columnwidth]{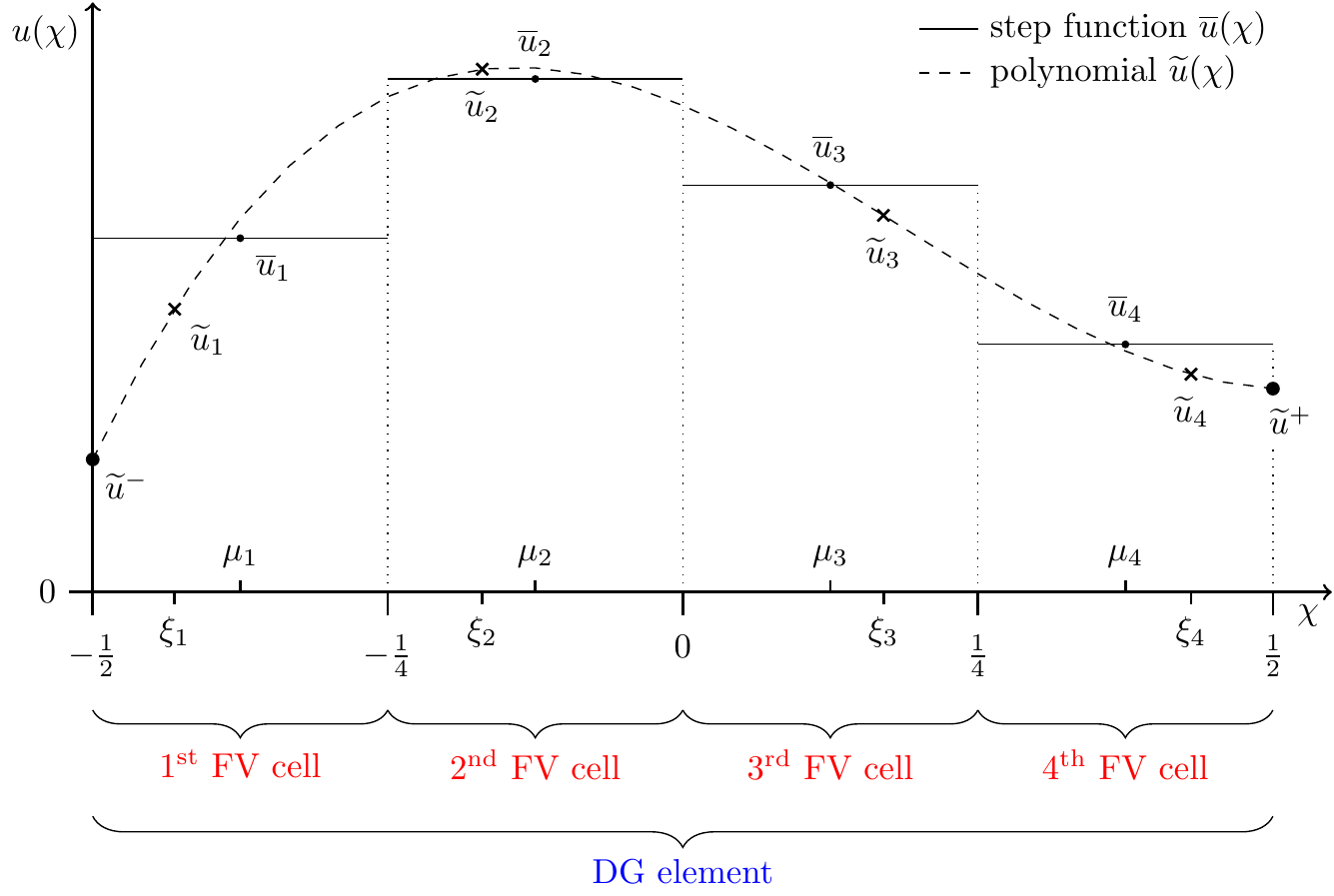}
\caption{1D schematic of four ($N = 4$) mean values $\overline{\svec{u}}_i$ and
their reconstructed nodal values (polynomial coefficients)
$\widetilde{\svec{u}}_i$ constituting a polynomial of degree $3$ spanning over
the whole DG element.}
\label{fig:reference-element}
\end{figure}

For shock capturing we now combine the robustness of FV and the accuracy of DG
via a convex blending scheme in the vicinity of discontinuous flow features.
In \citet{markert2021sub} this scheme is coined single-level blending scheme.
In the previous two subsections we have introduced the concepts of $N^3$ mean values
arranged within a block and the polynomial ansatz of $N^3$ nodal values living
inside an element. Now, we overlay the DG element with nodal values over the
block with mean values and get a new scheme that is a hybrid between FV and DG.
An illustration of this concept for a one-dimensional DG element and four mean
values ($N=4$) is shown in Fig.~\ref{fig:reference-element}.

\subsubsection{Projection \& Reconstruction}
\label{sec:projection-reconstructioin}
In order to make FV and DG compatible, we define projection and
reconstruction operators transforming between $N$ mean values
$\overline{\svec{u}}_i$ and $N$ nodal values $\widetilde{\svec{u}}_i$. For the
projection operator $\mathcal{P} \in \mathbb{R}^{N \times N}$, we make the
following ansatz with $i = 1,\dots,N$ 
\begin{equation}
\label{eq:def-projection}
\overline{\svec{u}}_i = N\int_{\mu_{i-\frac{1}{2}}}^{\mu_{i+\frac{1}{2}}} \widetilde{\svec{u}}(\chi) \, d\chi
    = \sum_{j=1}^N \widetilde{\svec{u}}_j \; \underbrace{N\,\int_{\mu_{i-\frac{1}{2}}}^{\mu_{i+\frac{1}{2}}} 
    \ell_j(\chi)\,d\chi}_{:= \; \mathcal{P}_{ij}},
\end{equation}
where the sub-cell centres, $\mu_i$, and interfaces, $\mu_{i\pm\frac{1}{2}}$, given as
\begin{equation}
\label{eq:subcell-grid}
    \mu_i = -\frac{1}{2} + \frac{i}{N} - \frac{1}{2\,N} \quad \text{and} \quad
    \mu_{i\pm\frac{1}{2}} = \mu_i \pm \frac{1}{2\,N},
\end{equation}
encode a regular sub-cell grid of $N$ finite volumes within the reference
element $\big[-\frac{1}{2},\frac{1}{2}\big]$.
By construction the projection matrix $\mathcal{P}$ is quadratic and
non-singular. Hence, the inverse $\mathcal{R} := \mathcal{P}^{-1}$ reconstructs
$N$ nodal values from given $N$ mean values. We write
\begin{align}
\label{eq:def-reconstruction-3d}
    &\overline{\svec{u}}_{ijk} = \sum_{c=1}^N \mathcal{P}_{kc} \sum_{b=1}^N \mathcal{P}_{jb} \sum_{a=1}^N \mathcal{P}_{ia} \:\:\widetilde{\svec{u}}_{abc} := \mathcal{P}_{ijk}^{[abc]} \:\:\widetilde{\svec{u}}_{[abc]}\quad \text{and} \quad \nonumber \\
    &\widetilde{\svec{u}}_{ijk} = \sum_{c=1}^N \mathcal{R}_{kc} \sum_{b=1}^N \mathcal{R}_{jb} \sum_{a=1}^N \mathcal{R}_{ia} \:\:\overline{\svec{u}}_{abc} := \mathcal{R}_{ijk}^{[abc]} \:\:\overline{\svec{u}}_{[abc]},
\end{align}
where we introduce a variant of the \textsc{Einstein} notation for brevity.
That is, indices enclosed in brackets are summed over.

\subsubsection{Convex Blending}
We aim to blend the right-hand-side solution of a LOW and HIGH
order scheme with a continuous blending factor $\alpha \in [0,1]$ in a convex manner:
\begin{equation*}
\partial_t \svec{u}_{q} \; = \; (1-\alpha)\,\partial_t \svec{u}_{q}^{\text{LOW}} \; + \; \alpha\,\partial_t \svec{u}_{q}^{\text{HIGH}}.
\end{equation*}
In order to make the disparate solution spaces for FV and DG compatible we
transform the nodal values from DG to mean values and convex combine with the FV
flux to the blended right-hand-side:
\begin{equation}
\label{eq:blending-scheme}
\partial_t \overline{\svec{u}}_{q,ijk} \; = \; (1-\alpha_{q})\,\partial_t \overline{\svec{u}}_{q,ijk}^{\text{FV}}
    \; + \; \alpha_{q}\,\mathcal{P}_{ijk}^{[abc]} \, \partial_t \widetilde{\svec{u}}_{q,[abc]}^{\text{DG}}.
\end{equation}
We call $\alpha_{q}$ the {\it volume blending factor}, which can be chosen
to be unique for each element. However, we have to carefully
ensure the conservation property of the blending scheme
by determining a common surface flux
\begin{equation}
\label{eq:common-surface-flux}
    \overline{\svec{F}}^{*}_{q\pm\frac{1}{2},jk} 
    = (1-\alpha_{q\pm\frac{1}{2}})\,\overline{\svec{F}}^{*\text{FV}}_{q\pm\frac{1}{2},jk} 
    + \alpha_{q\pm\frac{1}{2}}\,\mathcal{P}_{jk}^{[bc]}\,\widetilde{\svec{F}}^{*\text{DG}}_{q\pm\frac{1}{2},[cb]}.
\end{equation}
We call $\alpha_{q\pm\frac{1}{2}}$ the {\it surface blending factor}, which is
shared between neighbouring elements $q$ and $q~+~1$.
The outermost fluxes in \eqref{eq:3D-fv-scheme} are replaced with expression
\eqref{eq:common-surface-flux}:
\begin{equation*}
    \overline{\svec{F}}^{*}_{q,1jk} \rightarrow \overline{\svec{F}}^{*}_{q-\frac{1}{2},jk} \quad \text{and} \quad
    \overline{\svec{F}}^{*}_{q,Njk} \rightarrow \overline{\svec{F}}^{*}_{q+\frac{1}{2},jk}.
\end{equation*}
Likewise, we replace the surface flux in \eqref{eq:3D-dg-scheme} with the transformed
flux \eqref{eq:common-surface-flux}:
\begin{equation*}
    \widetilde{\svec{F}}^{*}_{q\pm\frac{1}{2},jk} \rightarrow \mathcal{R}_{jk}^{[bc]}\,\overline{\svec{F}}^{*}_{q\pm\frac{1}{2},[bc]}.
\end{equation*}
The surface fluxes in y- and z-direction are computed analogously.

\subsubsection{Calculation of the Blending Factor $\alpha$}
\label{sec:calc-blending-factors}
\begin{figure}
\centering
\includegraphics[width=1.0\columnwidth]{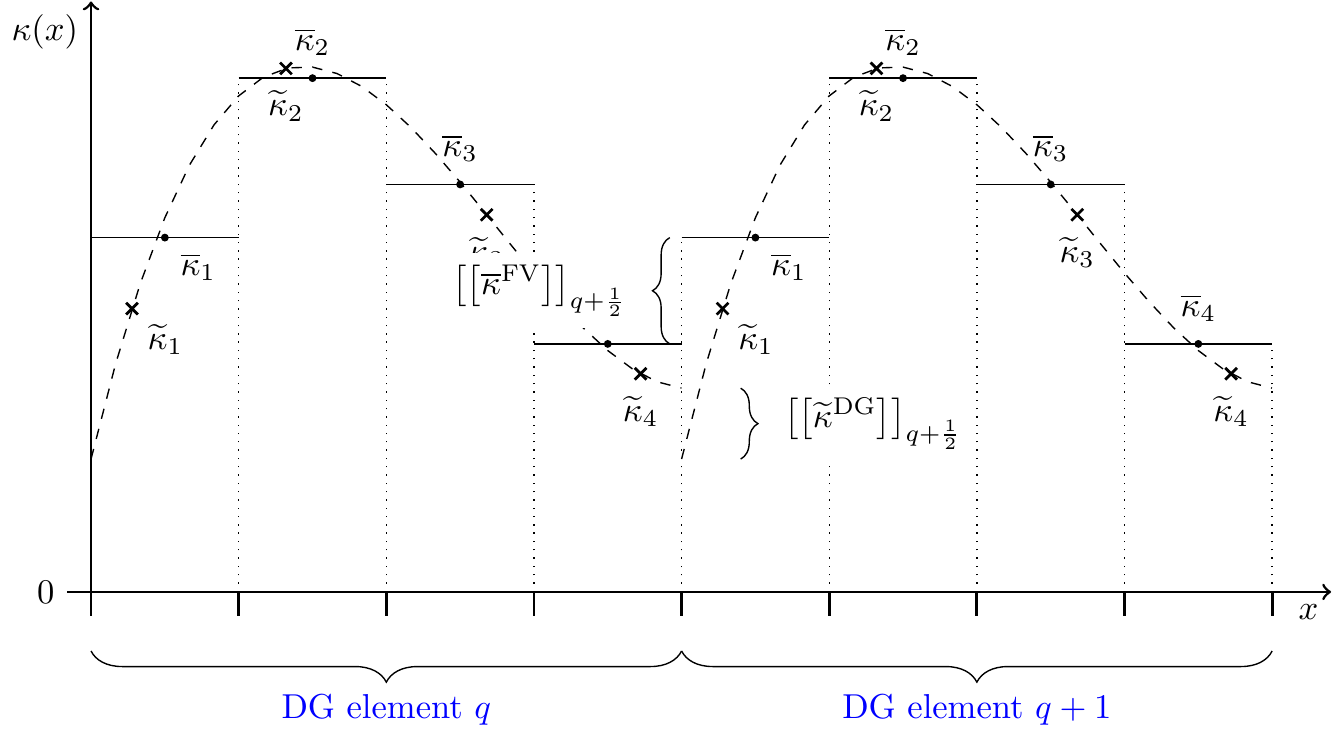}
\caption{1D schematic of two neighbouring elements $q$ and $q+1$ each with four
($N = 4$) mean values of the indicator variable $\overline{\kappa}_i$ and their
reconstructed polynomial coefficients $\widetilde{\kappa}_i$.  The relative
difference of the jumps at an element interface is considered to be a measure
of the smoothness of the solution at hand.}
\label{fig:block-jump}
\end{figure}

The surface blending factors $\alpha_{q\pm\frac{1}{2}}$ are estimated from the relative differences in the jumps of the element mean values
$\overline{\svec{u}}_{q,ijk}$ and the reconstructed polynomial
$\widetilde{\svec{u}}_{q,ijk}$ at element interfaces $q\pm\frac{1}{2}$. As in
the previous section, we transform the interpolated nodal interface values to
mean values beforehand:
\begin{equation*}
    \overline{\kappa}^{\text{DG}}_{q\pm\frac{1}{2},jk} = \mathcal{P}^{[bc]}_{jk}\;\widetilde{\kappa}_{q\pm\frac{1}{2},[bc]},
\end{equation*}
where $\kappa$ is a freely chosen {\it indicator variable}, such as density or pressure.
Additionally, we introduce the element interface jumps
\begin{equation}
\label{eq:block-interface-jump}
    \jump{\cdot}_{q\pm\frac{1}{2},jk} = (\cdot)^-_{q\pm\frac{1}{2},jk} - (\cdot)^+_{q\pm\frac{1}{2},jk}
\end{equation}
illustrated in Fig.~\ref{fig:block-jump}. The blending factor in
x-direction then reads
\begin{equation}
\label{eq:x-surface-blending-factor}
    \alpha_{q\pm\frac{1}{2},jk} = 1-\mathcal{T}\left(\jump{\overline{\kappa}^{\text{FV}}}_{q\pm\frac{1}{2},jk},\jump{\overline{\kappa}^{\text{DG}}}_{q\pm\frac{1}{2},jk}\right),
\end{equation}
with the transfer function
\begin{equation}
\label{eq:transfer-function}
\mathcal{T}\left((\cdot)^\text{FV},(\cdot)^\text{DG}\right) = \Bigg \lfloor_{0}
    \tau_A \, \frac{\left|\,(\cdot)^\text{FV}-(\cdot)^\text{DG}\right| - \tau_S\,\big|\,(\cdot)^\text{FV}\,\big|\,}
        {\max\big(\,\big|\,(\cdot)^\text{FV}\,\big|,1\,\big)}
    \Bigg\rceil_{1},
\end{equation}
mapping the two input arguments $(\cdot)^\text{FV},(\cdot)^\text{DG} \in
\mathbb{R}$ to the unit interval $[0,1]$ as is indicated by
the notation $\lceil_0\,(\cdot)\,\rceil_1$ trimming the input $(\cdot)$
for values below 0 and above 1. The {\it amplification parameter},
$\tau_A := 20$, and the {\it shifting parameter}, $\tau_S := 1.0$, are fixed for
all numerical results shown in this work. Moreover, we choose the
thermal pressure \eqref{eq:thermal-pressure} as the indicator variable.

For the common surface blending factor we pick the minimum result along the
interface:
\begin{equation}
\label{eq:surface-blending-factor}
    \alpha_{q\pm\frac{1}{2}} = \min_{ij=1}^N \alpha_{q\pm\frac{1}{2},jk}.
\end{equation}
For the volume blending factor $\alpha_{q}$ we calculate the averages of the
surface blending factors in each direction, for example in x-direction
\begin{equation*}
\label{eq:x-volume-blending-factor}
    \alpha^{x}_{q} = \frac{1}{2}\big(\alpha^{x}_{q-\frac{1}{2}} + \alpha^{x}_{q+\frac{1}{2}}\big),
\end{equation*}
and determine the minimum among all directions
\begin{equation}
\label{eq:volume-blending-factor}
    \alpha_{q} = \min\big\{\alpha^{x}_{q},\alpha^{y}_{q},\alpha^{z}_{q}\big\}.
\end{equation}

The idea of the proposed algorithm is to have a mechanism which is designed
such that for well resolved flows the solver yields the full DG solution and
gradually shifts to the FV solution in case of discontinuities or strong
under-resolution. Additionally, the blending factor is set to zero, if the
reconstructed polynomial coefficients yield unbound values, such as
negative densities or negative pressures. The result is then a 100\%
second-order slope limiting FV solution in elements where the reconstruction of
a DG polynomial failed. Hence, in some sense, the proposed method with convex
blending of a low and high-order operator guarantees that the discretisation
cannot be "worse" than the second-order FV method. 

\subsection{Source terms}
\label{sec:source-terms-disc}
The discretisation of the gravity source term \eqref{eq:gravity-source-term} is
constructed in a straight forward manner by inserting the cell-averaged fluid state, i.e.
\begin{equation*}
\overline{\svec{\Upsilon}}^{\text{gravity}}_{q,ijk} =
\svec{\Upsilon}^{\text{gravity}}\big(\overline{\svec{u}}_{q,ijk}\big),
\end{equation*}
where the gravitational acceleration vector $\vec{g}$ in \eqref{eq:gravity-source-term} comes in the form of mean values
from the gravity modules provided by FLASH \citep[e.g. the tree-based gravity
solver by][]{wunsch2018tree}. We do remark that this way of evaluating the
source term, albeit very robust and straight forward to implement, is formally
only second order accurate as the mean value of the sub-cell and not the DG
polynomial evaluation is used. However, as already mentioned, the
goal is to connect to the multi-physics framework as seamless as possible and
this is one compromise taken.

\subsection{Time Integration}
Time integration of the semi-discrete equation \eqref{eq:3D-fv-scheme} is done
by explicit Runge-Kutta schemes, such as the optimal second order Ralston's Method
\citep{ralston1962runge} or the fourth order, five stages and strong-stability
SSP-RK(5,4) variant \citep{spiteri2002new}.

To calculate a stable timestep the maximum eigenvalue estimate
\eqref{eq:mhd-max-wave-speed} is evaluated on all mean values
$\overline{\svec{u}}_{q,ijk}$ in block $q$. In three dimensions it reads as
\begin{equation}
\label{eq:euler-max-eigenvalue}
\overline{\lambda}_{q}^{\text{max}} = \max_{d=1}^{3}\max_{ijk=1}^{N} \lambda^{\text{max}}_d\Big(\overline{\svec{u}}_{q,ijk}\Big).
\end{equation}
Then the maximal timestep is estimated by the CFL condition
\begin{equation}
\label{eq:cfl-condition}
    \Delta t := \frac{CFL}{d} \, \min_q\;\frac{\min(\Delta x_q,\Delta y_q,\Delta z_q)}{N\;\overline{\lambda}_{q}^{\text{max}}},
\end{equation}
where $d = 3, CFL := 0.4$ and $N$ is the number of mean values in each
direction of the block $q$.  Furthermore, we calculate the global hyperbolic
correction speed
\eqref{eq:global-hyperbolic-correction-speed} with
\begin{equation}
\label{eq:glm-correction-speed}
    c_h = \max_q\,\max_{d=1}^{3}\max_{ijk=1}^{N} c_d\Big(\overline{\svec{u}}_{q,ijk}\Big).
\end{equation}

\subsection{Enforcing Density and Pressure Positivity}
In the conserved variables formulation \eqref{eq:balance-law}, the thermal
pressure \eqref{eq:thermal-pressure} is derived from the total energy by
subtracting the kinetic and magnetic energy terms. Hence, at strong shocks or
under near-vacuum conditions the scheme can produce non-physical states. To
alleviate this problem, we lift the troubled cells into positivity such that
the permissibility condition 
\begin{equation}
\label{eq:permissible-states}
    \big\{\text{permissible states}\big\} = \big\{\forall\,\svec{u} \; \big| \; \rho > 0 \wedge p(\svec{u}) > 0\big\}.
\end{equation}
is fulfilled for all cells in a block $q$.  First, we calculate the block
average
\begin{equation}
\label{eq:block-average}
    \langle\svec{u}\rangle_{q} = \frac{1}{N^3}  \sum_{ijk=1}^N \, \overline{\svec{u}}_{q,ijk}
\end{equation}
and then determine a {\it 'squeezing' parameter} $\beta \in [0,1]$ which enforces
physical states:
\begin{equation}
%\label{eq:block-average}
    \overline{\svec{u}}_{q,ijk}^{\text{physical}} = (1-\beta)\,\langle\svec{u}\rangle_{q}  + \beta\;\overline{\svec{u}}_{q,ijk} \quad \in \big\{\text{permissible states}\big\}.
\end{equation}
The algorithm to find a suitable $\beta$ is straightforward. One starts with
$\beta := 1$ and decrements in steps of $\Delta\beta := 0.1$ till the
permissibility condition is fulfilled. The advantage of this algorithm lies in its
simplicity and it is conservative by construction. However, it fails when the
block average is not part of the permissible set. In this case the code crashes
and the simulation stops. For the results presented in this paper this worst
case never happened.

\section{Implementation in FLASH}
\label{sec:implementation-in-flash}
The FLASH code is a modular, parallel multi-physics simulation code capable of
handling general compressible flow problems found in many astrophysical
environments. 
It uses the Message Passing Interface (MPI) library for inter-processor
communication and the HDF5 or Parallel-NetCDF library for parallel I/O to
achieve portability and scalability on a variety of different parallel
computers. The framework provides three interchangeable grid modules: a Uniform
Grid, a block-structured octree based adaptive grid using the PARAMESH library
\citep{olson1999PARAMESH} and a block-structured patch based adaptive grid using
Chombo \citep{colella2009chombo}. 
The architecture of the code is designed to be flexible and easily extensible.

The numerical scheme described in Section~\ref{sec:numerical-discretization}
has been implemented within FLASH as an independent solver module named
DGFV under the \textsc{hydro/HydroMain/split} namespace. Although our
numerical scheme is technically an {\it unsplit} solver it mimics the same
interface as the other {\it directionally split} solvers since
most physics modules necessary for our envisaged astrophysics simulations only
interface to directionally split schemes. Contrary to the described 3D FV scheme in
\Secref{sec:finite-volume-scheme} we cannot split the 3D DGSEM scheme from
\Secref{sec:dgsem} into separate arrays of 1D data; at least not without
compromising the accuracy. These so called {\it sweeps} in spatial direction
$d$ are treated independently by directionally split schemes and assembled to the new
solution afterwards. Fortunately, the split scheme interface is flexible enough
to accommodate a high-order (unsplit) DG scheme.

\begin{figure}
\centering
\includegraphics[width=0.9\columnwidth]{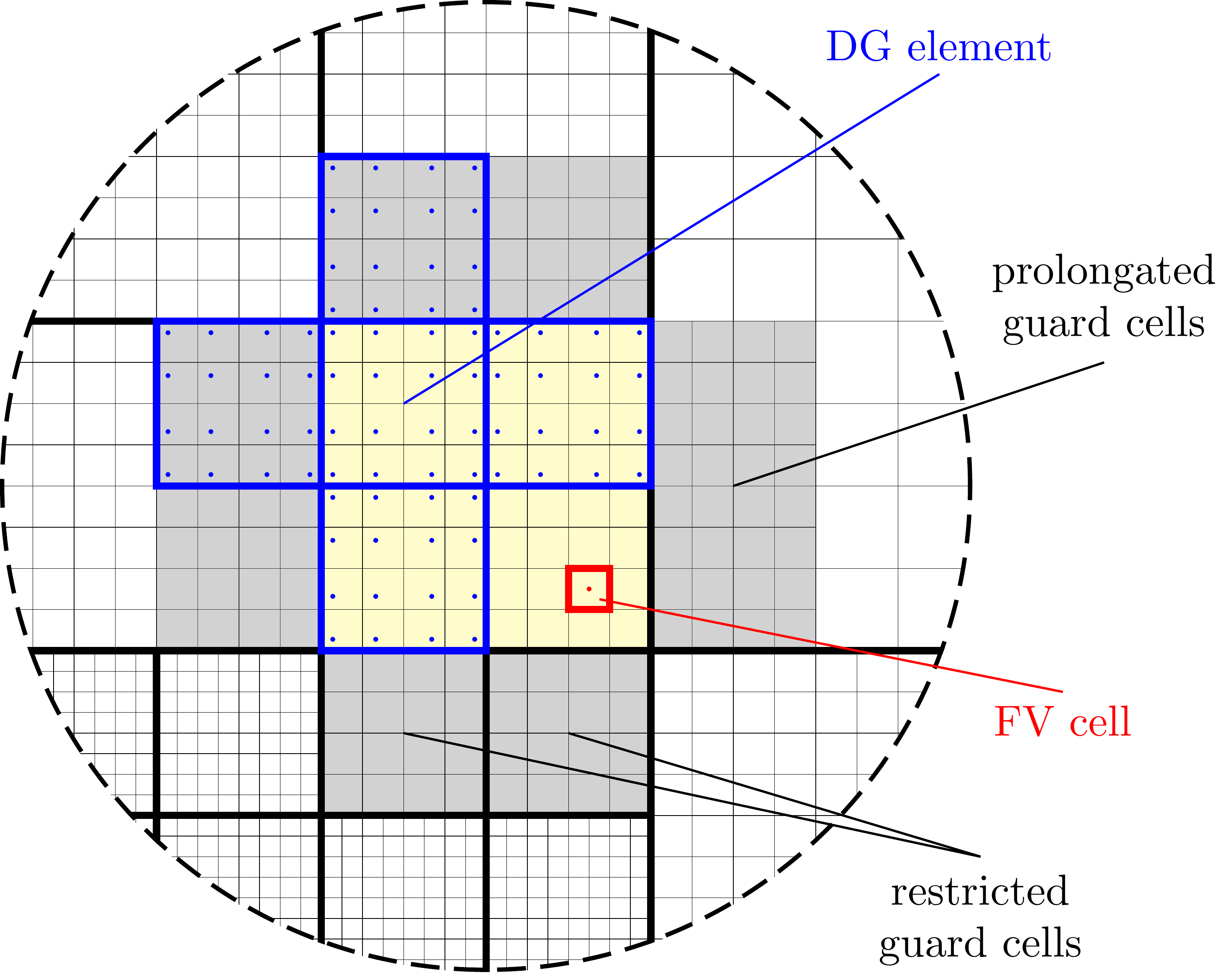}
\caption{Cross section of a PARAMESH grid with blocks of $8^3$ cells at
different refinement levels (i.e.  mesh resolution). The block of
interest is shaded in yellow and its guard cell layer is shaded in
grey. One 4$^{\rm th}$-order DG element with its adjacent elements is highlighted in blue and the blue dots indicate the position of the Gauss quadrature nodes \eqref{eq:quadrature-nodes}. The red square with its red midpoint exemplifies one FV cell in the block.}
\label{fig:PARAMESH2d}
\end{figure}
We focus our solver implementation on the octree-based grid unit
PARAMESH which is the default in FLASH. PARAMESH is built on the
basic component of blocks consisting of $N_x \times N_y \times N_z$ regular
cells extended with $N_g$ layers of guard cells.  Fig.~\ref{fig:PARAMESH2d}
shows a cross section of such a PARAMESH grid. The block of interest is coloured
in yellow while the guard cells are shaded in grey.  The complete computational
grid consists of a collection of such blocks with different physical cell
sizes. They are related to each other in a hierarchical fashion using a tree
data structure (octree).
Blocks do not overlap and there are only jumps of maximal one refinement level allowed between neighbours sharing a common face. 

PARAMESH handles the filling of the
guard cells with information from neighbouring blocks or at the boundaries of
the physical domain. If the block's neighbour has the same level of refinement,
PARAMESH fills the corresponding guard cells using a direct copy from the
neighbour's interior cells. If the neighbour is at a higher refinement level, the
data is {\it restricted} via second order averaging. If the neighbour is at a
lower refinement level, the data is {\it prolongated} via monotonic
interpolation guaranteeing positive densities and pressures.

The default configuration in FLASH is $N_x = N_y = N_z := 8$ and $N_g := 4$,
which allows us to implement a fourth order DG method ($N=4$) with $2^3$
elements embedded within a block. In Fig.~\ref{fig:PARAMESH2d} one DG element and
its DG neighbours are highlighted in blue. The blue dots represent the
\textsc{Legendre-Gauss} quadrature nodes \eqref{eq:quadrature-nodes}. The nodal state
values of the DG element are reconstructed from the underlying mean values as
described in Section~\ref{sec:projection-reconstructioin}.

What follows is a brief outline of the algorithm implemented within the DGFV
solver module. The module enters the Runge-Kutta cycle of $m$ stages starting
with the first stage. A full timestep is completed after exiting the loop with
the last stage where we compute the global timestep of the next cycle according
to the CFL condition \eqref{eq:cfl-condition}. The updated states are returned
back to FLASH, which in turn calls the other modules handling different aspects
of the simulation such as grid coarsening/refinement, load balancing, gravity,
chemistry, etc.  At the beginning of each Runge-Kutta stage the guard cells of
each block are filled by PARAMESH with the latest data from direct neighbours.
We reconstruct the nodal data of the DG elements via
\eqref{eq:def-reconstruction-3d} and compute the blending factors according to
Section~\ref{sec:calc-blending-factors}. If any of the calculated blending
factors is less than one, indicating under-resolved flow features within a DG
element, the blending procedures are activated and we compute the standard FV
solution described in Section~\ref{sec:finite-volume-scheme}. Otherwise it
suffices to just compute the DG solution from the reconstructed nodal values
according to Section~\ref{sec:dgsem}. If blending is active one has to
determine the common surface fluxes \eqref{eq:common-surface-flux} at DG
element interfaces in order to maintain the conservation property. The final
solution is then the convex blend \eqref{eq:blending-scheme} of both solutions.
Moreover, the (common) surface fluxes at block boundaries are handed over to
PARAMESH, which does a {\it flux correction} by replacing the coarse fluxes with
the restricted fine fluxes at non-conforming block interfaces. We retrieve the
corrected fluxes and calculate the total surface flux error among all block
boundaries. The error is then evenly deducted from all cells restoring mass
conservation. Finally, we calculate the gravity source terms
\eqref{eq:gravity-source-term}, which we add to the solution. The solver can now
proceed to the next Runge-Kutta stage.  This completes the outline of the
algorithm.

Three remarks are in order. Firstly, FLASH stores its fluid data in primitive
state variables. For DG we need to reconstruct on conservative state variables
in order to maintain high-order accuracy and stability. In other words, the
conservative state variables are first calculated from the primitive state
variables in mean value space and then transformed to nodal values. Secondly,
the treatment at non-conforming block interfaces by PARAMESH is at most second
order accurate. For DG the standard approach is the so called {\it mortar
method} \citep{Koprivaetal2000}, which matches the spatial order of the scheme if
done correctly. However, we decided to stick with the default procedure
provided by PARAMESH since it would otherwise lead to substantial restructuring
of the internal workings of the FLASH code. Furthermore, we observed that in
practice the gain in accuracy by a higher-order mortar method is negligible in
our simulations, neither do we have any troubles with numerical artefacts or
stability issues. Thirdly, physics and chemistry units in FLASH are designed
and implemented around mean values, i.e. they expect mean values as input, do
their calculations on mean values and produce mean values as output. As
mentioned already in \Secref{sec:source-terms-disc} about source terms,
this \textit{formally} reduces the convergence order to at most second order, however,
provides the benefit of directly using the rich collection of physics modules
available in the FLASH framework.

\section{Numerical Results}
In this section, we present simulation results using our new 4$^{\rm th}$-order
(DGFV4) fluid solver module in FLASH. All test problems shown are computed on
1D, 2D or 3D Cartesian grids, for which the refinement level $l$ corresponds to
$2^{l-1}\times8$ grid points per dimension. We gradually increase the
complexity of the test cases, where the later test cases are multi-physics
applications with multiple physics modules working together.

\subsection{Experimental Order of Convergence}
First, we verify the experimental order of convergence (EOC) of our 4$^{\rm
th}$-order DGFV4 implementation. Measuring convergence of higher-order MHD
codes can be challenging since smooth MHD test problems present numerical
subtleties that are potentially revealed by the very low numerical dissipation
of higher-order methods. In this work, we rely on common, smooth MHD test
problems for convergence assessment: the non-linear circularly polarised Alfvén
waves in 2D and a manufactured solution setup for 3D.

For our convergence tests, we enable any shock indicators and limiters on
purpose in order to confirm that the limiters do not interfere for smooth and well
resolved solutions. 

To study the convergence properties of the code, we compute the numerical
errors, that is, the norm of the difference of the numerical solution,
$u_{\text{num}}$, from the reference solution, $u_{\text{ref}}$, with mean values.
In this work, we look at the $\infty$-norm, i.e.
\begin{equation}
\label{eq:infty-norm}
\big|\big|\overline{u}_{\text{num}} - \overline{u}_{\text{ref}}\big|\big|_{\infty}
    = \max_{q=1}^Q \max_{ijk=1}^N \Big|\overline{u}_{\text{num},q,ijk} - \overline{u}_{\text{ref},q,ijk}\Big|
\end{equation}
and the 2-norm, i.e.
\begin{equation}
\label{eq:2-norm}
\big|\big|\overline{u}_{\text{num}} - \overline{u}_{\text{ref}}\big|\big|_2 
    = \left[|\Omega|^{-1} \sum_{q=1}^Q \sum_{ijk=1}^N \frac{V_{q}}{N^3}\Big|\overline{u}_{\text{num},q,ijk} - \overline{u}_{\text{ref},q,ijk}\Big|^2\right]^{\frac{1}{2}},
\end{equation}
where $V_{q} = \Delta x_q\,\Delta y_q\,\Delta z_q$ is the volume of block $q$
and $|\Omega|$ is the volume of the computational domain.

We note that one has to compute the mean values of the reference solution with
high accuracy as they should formally be the exact mean values of the reference
solution. Hence, we evaluate the reference solution on quadrature nodes of the
same order or higher as in \eqref{eq:volume-flux} and then project to mean
values.

Circularly polarised smooth Alfvén waves are exact analytic solutions of the
MHD equations for arbitrary wave amplitude perturbations. Hence, they are
suitable to study the experimental convergence order of the scheme, as well as
its amount of numerical dispersion and dissipation errors. The solution
consists of plane waves in which the magnetic field and velocity oscillate in
phase in a circular polarisation perpendicular to the propagation direction. 

We initialise the Alfvén waves within a periodic 2D domain $\Omega = [0,0]
\times [\cos^{-1}(\alpha),\sin^{-1}(\alpha)]$ at an angle of $\alpha =
45^{\circ}$.  The longitudinal Alfvén wave speed is fixed at $|v_A| =
B_{||}/\sqrt{\rho} := 1$, such that the wave returns to its initial state at
integer times $T \in \mathbb{N}$. For our test we let the wave turn around five
times, i.e.  $T = 5$. The heat capacity ratio is $\gamma = 5/3$.  With the
rotated coordinate $\hat{x} = x\,\cos(\alpha) + y\,\sin(\alpha)$, we define
$B_{\perp} = 0.1\,\sin(2\pi\,\hat{x})$ and $B_3 = 0.1\,\cos(2\pi\,\hat{x})$.
The initial state in primitive state variables then reads
$(\rho,\vec{v},p,\vec{B},\Psi)^T = (1.0,\vec{v}_0,0.1,\vec{B}_0,0)^T$ with
$\vec{v}_0 = (-B_{\perp}\sin(\alpha),B_{\perp}\cos(\alpha),B_3)^T$ and
$\vec{B}_0 = (\cos(\alpha),\sin(\alpha),0)^T + \vec{v}_0$. We look at the total
pressure \eqref{eq:total-pressure} and calculate the $\infty$-norm and 2-norm
according to \eqref{eq:infty-norm} and \eqref{eq:2-norm}. Since the total
pressure involves all state variables it is a good quantity to measure the
overall convergence of the code. The results are shown in
Table~\ref{tab:eoc-alfven-wave-dgfv4}. The formally 4$^{\rm th}$-order DGFV4
scheme in 2D yields the expected EOCs.
\begin{table}
\centering
\caption{EOC of total pressure $P$ of the smooth Alfvén wave problem in 2D run
by the 4$^{\rm th}$-order DGFV4 scheme.}
\label{tab:eoc-alfven-wave-dgfv4}
\begin{tabular}{r|cc|cc}
\hline
cells & $\big|\big|P\big|\big|_{\infty}$ & $\big|\big|P\big|\big|_{2}$ & EOC$_{\infty}$ & EOC$_{2}$  \\
\hline
$16^2$  &  9.852e-05 &  4.951e-05  &    n/a   &  n/a  \\
$32^2$  &  6.086e-06 &  2.110e-06  &  4.016   & 4.552 \\
$64^2$  &  3.634e-07 &  1.285e-07  &  4.065   & 4.036 \\
$128^2$ &  2.334e-08 &  8.002e-09  &  3.960   & 4.006 \\
$256^2$ &  1.461e-09 &  4.998e-10  &  3.997   & 4.000 \\
$512^2$ &  9.221e-11 &  3.423e-11  &  3.985   & 3.868 \\
\hline
\end{tabular}
\end{table}

Next, in order to verify the convergence order in 3D we construct a manufactured
solution which, in primitive state variables, reads as
\begin{equation}
\label{eq:manufac-solution}
\svec{\Phi}^{\text{man}} = \big(h,0.1,0.2,0.3,h,h,-h,1,0\big)^T
\end{equation}
with $h = h(t,x,y,z) = 0.5\sin(2\pi(x + y + z - t))$. The domain is a cubic box
$\Omega = [0,1]^3$ and the heat capacity ratio is set to $\gamma = 2.0$. At
each Runge-Kutta stage we subtract the residual
\begin{equation}
%\label{eq:manufac-solution}
\svec{\Upsilon}^{\text{man}} 
    = \partial_x F\big|_{\svec{\Phi}^{\text{man}}}
    + \partial_y G\big|_{\svec{\Phi}^{\text{man}}}
    + \partial_z H\big|_{\svec{\Phi}^{\text{man}}}
\end{equation}
from the numerical solution $\partial_t\svec{u}$ and advance in time. As before,
the source term is evaluated with an appropriate quadrature rule and projected
to mean values. At final time $T = 0.1$ the convergence of the total pressure \eqref{eq:total-pressure} is computed as previously described in the Alfvén wave test. The results in Table~\ref{tab:eoc-manufac-dgfv4} confirm the correct EOCs for the formally
4$^{\rm th}$-order DGFV4 scheme in 3D.
\begin{table}
\centering
\caption{EOC of total pressure $P$ of the manufactured solution problem in 3D run
by the 4$^{\rm th}$-order DGFV4 scheme.}
\label{tab:eoc-manufac-dgfv4}
\begin{tabular}{r|cc|cc}
\hline
cells & $\big|\big|P\big|\big|_{\infty}$ & $\big|\big|P\big|\big|_{2}$ & EOC$_{\infty}$ & EOC$_{2}$  \\
\hline
$16^3$  &  1.259e-03 &  4.122e-04  &  n/a   &   n/a  \\
$32^3$  &  4.998e-05 &  1.773e-05  &  4.655 &  4.538 \\
$64^3$  &  3.248e-06 &  1.015e-06  &  3.943 &  4.126 \\
$128^3$ &  1.962e-07 &  6.594e-08  &  4.048 &  3.944 \\
$256^3$ &  1.290e-08 &  4.141e-09  &  3.927 &  3.992 \\
$512^3$ &  7.963e-10 &  2.540e-10  &  4.018 &  4.027 \\
\hline
\end{tabular}
\end{table}

\subsection{Divergence Control}
We now turn to test problems more specifically aimed at evaluating the efficacy
of the hyperbolic divergence cleaning scheme discussed in
\Secref{sec:hyperbolic-div-cleaning}.

\subsubsection{Magnetic Loop Advection}
This test investigates the proper advection of a magnetic field loop
\citep{gardiner2005unsplit}. On a periodic domain $\Omega = [-0.5,0.5]^2$, the
background medium has $\rho = 1$, $p = 1$, and a global advection velocity
$(v_1 v_2) = (2, 1)$ so that the ambient flow is not aligned with any
coordinate axis. Letting $r = \sqrt{x^2+y^2}$ be the radial distance to the
centre of the domain, the magnetic field is initialised from a vector potential
$\vec{A} = (0, 0, A_3(r))$ with $\vec{B} = \nabla \times \vec{A}$. To define a
magnetic field loop of radius $r_0 = 0.3$, we set $A_3(r) = \max(0,10^{-3}(r_0
- r))$ and obtain a very weakly magnetised configuration with a plasma $\beta$
  of order $10^6$, in which the magnetic field is essentially a passive scalar.
For this field configuration, the MHD current vanishes everywhere, except at $r
= 0$, and $r = r_0$ where the corresponding current line and current tube
become singular. Note, that we added a smoothing parameter $\delta r = 0.05$ to
$r$ in order to avoid spurious ringing caused by the unresolved singularity of
the initial magnetic field at the domain centre. The aim of the test is to
verify that the current loop is advected without deformation or noise.

\begin{figure}
\centering
\includegraphics[width=1.0\columnwidth]{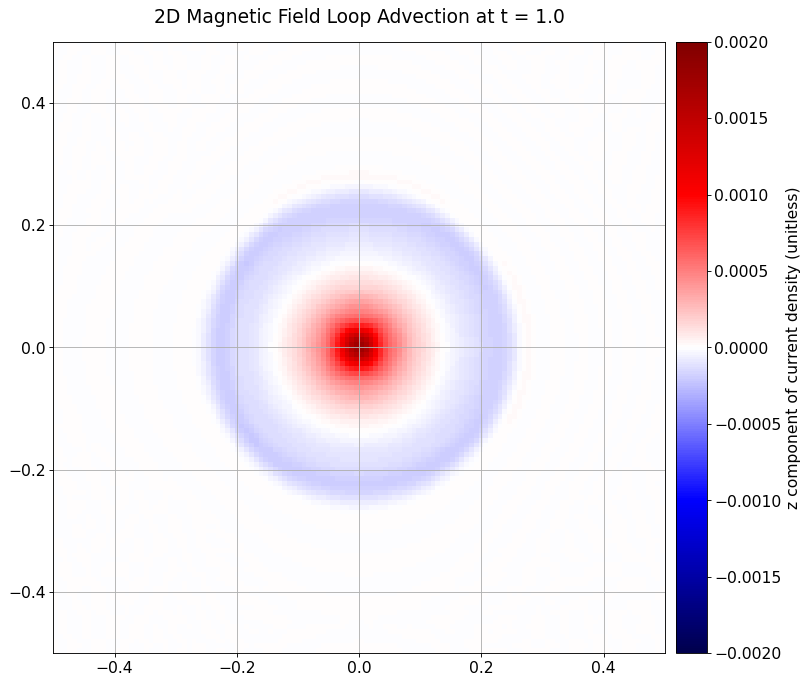}
\caption{We show the $z$-component of the current density $\vec{j} = \nabla \times \vec{B}$
for the final time $T = 1.0$ as calculated with the DGFV4 scheme.}
\label{fig:loop-current-density}
\end{figure}

\Figref{fig:loop-current-density} shows the z-component of the current
density $\vec{j} = \nabla \times \vec{B}$ at final time $T = 1$ corresponding
to two horizontal domain crossings. The test is carried out with a uniform
resolution of $128^2$ cells.  The current density is a stringent diagnostic
since, being a derivative of the magnetic field, it is very sensitive to noise
and local fluctuations. The scheme preserves the exact circular shape of the
current loop very well, with very little noise and oscillations in the current.

\subsubsection{MHD Current Sheet}
\label{sec:current-sheet}
The two-dimensional current sheet problem in ideal MHD regimes has been
extensively studied before \citep[see
e.g.][]{gardiner2005unsplit,guillet2019high,rastatter1994current,rueda2021entropy}.
Two magnetic currents are initialized in opposite direction sharing a common
interface and disturbed with a small velocity fluctuation which provokes
magnetic reconnections. In the regions where the magnetic reconnection takes
place, the magnetic flux approaches very small values and the lost magnetic
energy is converted into internal energy. This phenomenon changes the overall
topology of the magnetic fields and consequently affects the global magnetic
configuration.

The square computational domain is given as $\Omega = [-0.5,0.5]^2$ with
periodic boundary conditions. We initialise the setup in primitive variables
$(\rho,\vec{v},p,\vec{B},\Psi)^T = (1.0, v_1,0,0, 0.05\,B_0^2, 0,B_2.0,0)^T$
with $B_0 = 1/\sqrt{4\pi}$, $v_1 = 0.1\sin(2\pi y)$ and $B_2 = -B_0$
when $-0.25 < x < 0.25$ and $B_2 = B_0$ in the rest of the domain.  The heat
capacity ratio is set to $\gamma = 5/3$ and the final simulation time is $T =
10$. The simulation is carried out with a uniform
resolution of $256^2$ cells.

\begin{figure}
\centering
\includegraphics[width=1.0\columnwidth]{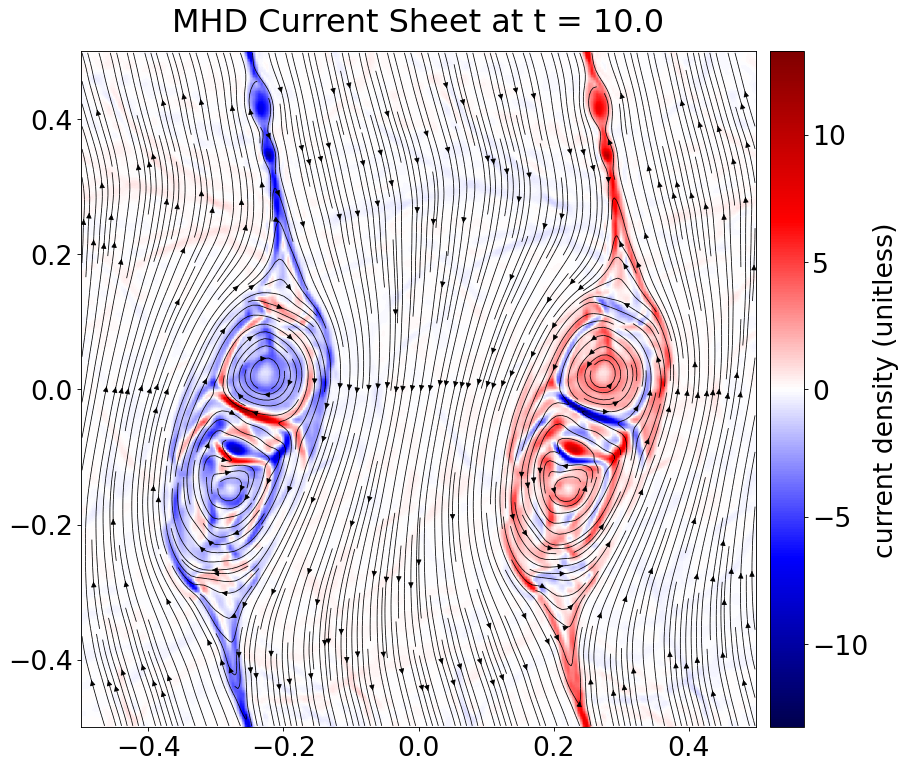}
\caption{We show the $z$-component of the current density $\vec{j} = \nabla \times \vec{B}$
with overlayed magnetic field lines for the MHD current sheet test at final time
$T = 10$.}
\label{fig:current-sheet-current-density}
\end{figure}

The changes in the magnetic field seed the magnetic reconnection and develop
formations of magnetic islands along the two current sheets. The small islands
are then merged into bigger islands by continuously shifting up and down along
the current sheets until there is one big island left in each current sheet.
The final result is shown in \Figref{fig:current-sheet-current-density}, which
depicts the $z$-component of the current density with overlayed magnetic field
lines.

\begin{figure}
\centering
\includegraphics[width=1.0\columnwidth]{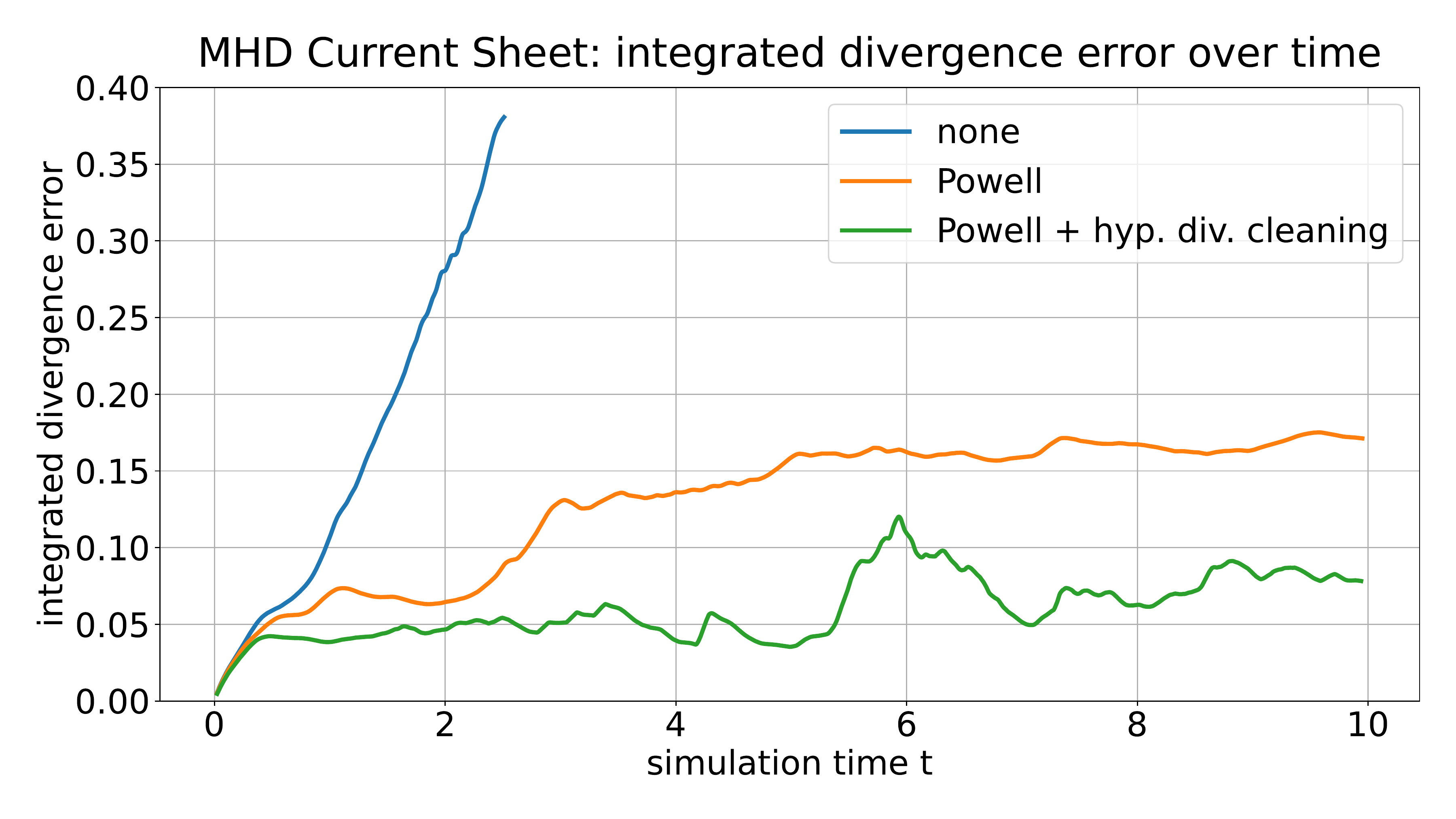}
\caption{
Evolution of the domain-integrated absolute magnetic field divergence $\int_{\Omega} |\nabla\cdot\vec{B}(t)| d\Omega$ for the MHD current sheet test. Three runs with different flavours of divergence cleaning are shown. The simulation without any
divergence correction method ("none"; blue line) crashed at around $t = 2.5$.
The run where only the Powell source terms are used (orange line) is stable but
has a substantially larger divergence error than the full scheme with Powell
source terms and hyperbolic divergence cleaning (green line) which is the
default setting in our code.}
\label{fig:current-sheet-divergence-over-time}
\end{figure}

\begin{figure}
\centering
\includegraphics[width=1.0\columnwidth]{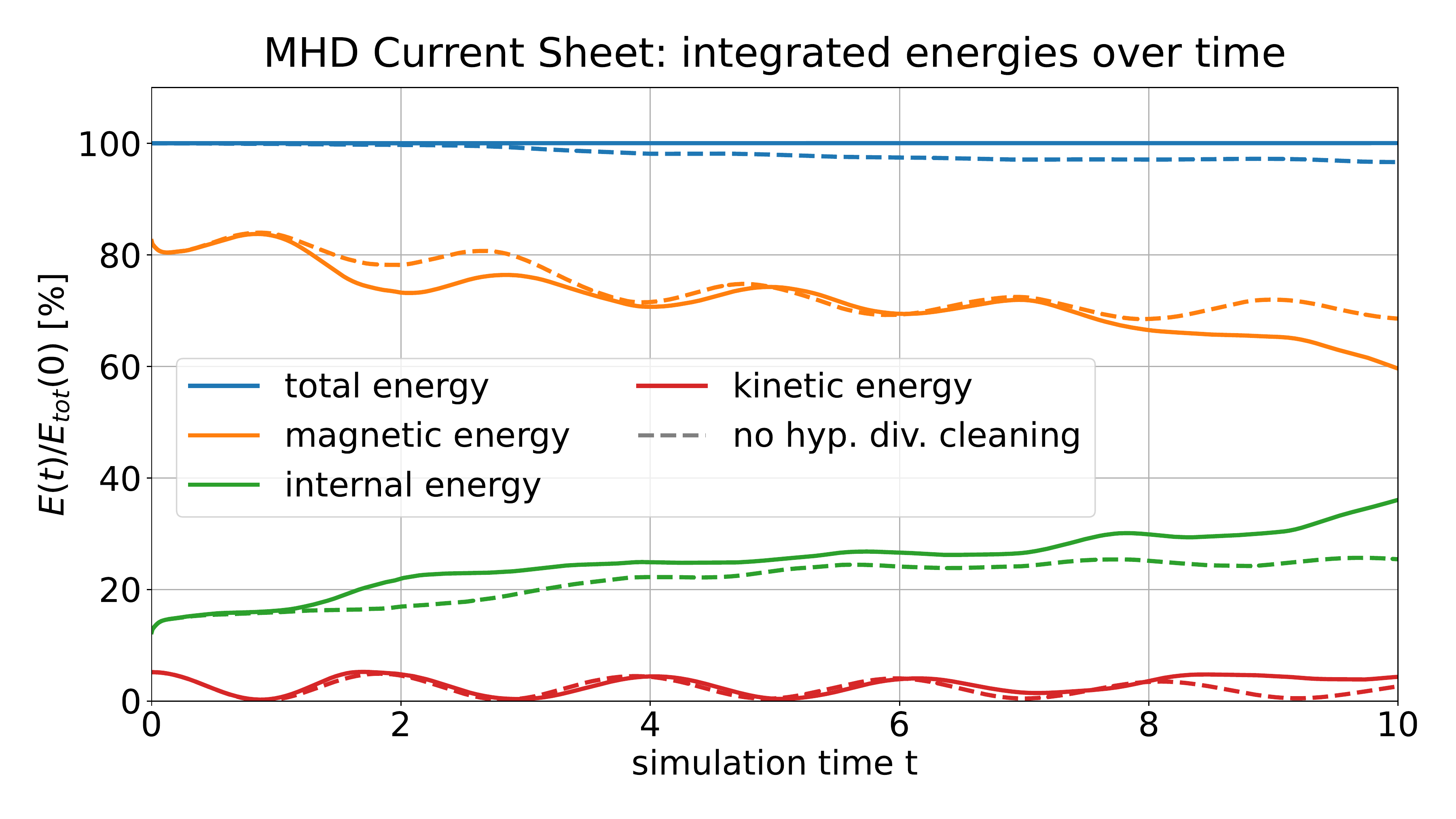}
\caption{The percentage of total and individual energies as a function of time
for the MHD current sheet test. We compare the results with (solid lines) and
without (dashed lines) our hyperbolic divergence cleaning method. The results for
the run without any divergence treatment is not shown since it crashed early
on.}
\label{fig:current-sheet-energies-over-time}
\end{figure}

\Figref{fig:current-sheet-divergence-over-time} shows the domain-integrated divergence
error over time for three different runs: without any divergence cleaning
(none; blue line), with just Powell source terms (Powell; orange line) and with
Powell source terms plus hyperbolic divergence cleaning (Powell + hyb. div.
cleaning; green line) which is the default in our code. The first simulation crashes early on due to the rapid
surge of divergence error leading to the well known instability issues of
uncontrolled magnetic field divergence growth in MHD simulations
\citep{brackbill1980effect,toth2000b,kemm2013origin}.  The second run with
Powell terms does not crash and is at least able to keep the overall errors at
bay over the course of the simulation (see also
Fig.~\ref{fig:current-sheet-energies-over-time}). As expected, the run with
hyperbolic divergence cleaning has the lowest divergence error at all times.
This particular setup is insofar challenging in that it perpetually prompts a
significant production of divergence errors, thus highlighting the importance
of a proper and robust divergence cleaning technique.

\Figref{fig:current-sheet-energies-over-time} shows the time evolution of the
kinetic, internal, magnetic, and total energies of the current sheet problem
with (solid lines) and without (dashed lines) hyperbolic divergence cleaning
activated (see \Secref{sec:hyperbolic-div-cleaning}). The decline of the
magnetic energy is compensated by the increase in internal energy due to the
heating driven by the magnetic reconnection. Furthermore, without divergence
cleaning a significant total energy drift is introduced by the non-conservative
Powell terms and the accompanied high divergence errors, which are not properly
corrected for in this case. This energy drift also causes artificial cooling at
the reconnection points giving rise to a non-physical behaviour. However, with
hyperbolic divergence cleaning the total energy is conserved.

\subsection{Shock Problems}
In this section, we test the correct resolution of shock waves inherent to ideal
MHD regimes by our DGFV4 scheme in 1D, 2D and 3D. Furthermore, one shock tube
problem investigates the proper handling of waves in a multi-species setting
and we also utilise adaptive mesh refinement for a further challenge.
Note that we actually run all our 1D test problems in 2D, with perfectly
$y$-independent initial conditions and periodic boundaries. A correct
implementation of our multidimensional scheme does not develop any 
$y$-dependence
of the solution.

\subsubsection{Briu-Wu Shock Tube}
The established MHD shock tube problem introduced by \citet{brio1988upwind} has
now become a classic shock test for MHD codes. For this test, we take the
computational domain to be $[0,1]$ with outflow boundaries left and right. In
the whole domain, the flow is initially at rest ($v = 0$) and $(B_1,B_2) =
(0.75, 0)$. The initial primitive variables are discontinuous at $x = 0.5$,
with the left and right states given by $(\rho,p, B_2)_{\mathrm{L}} = (1, 1, 1)$ and
$(\rho,p, B_2)_{\mathrm{R}} = (0.125, 0.1, -1)$, respectively. We set $\gamma = 2$, and run
the simulation until the final time $T = 0.1$.

\begin{figure}
\centering
\includegraphics[width=1.0\columnwidth]{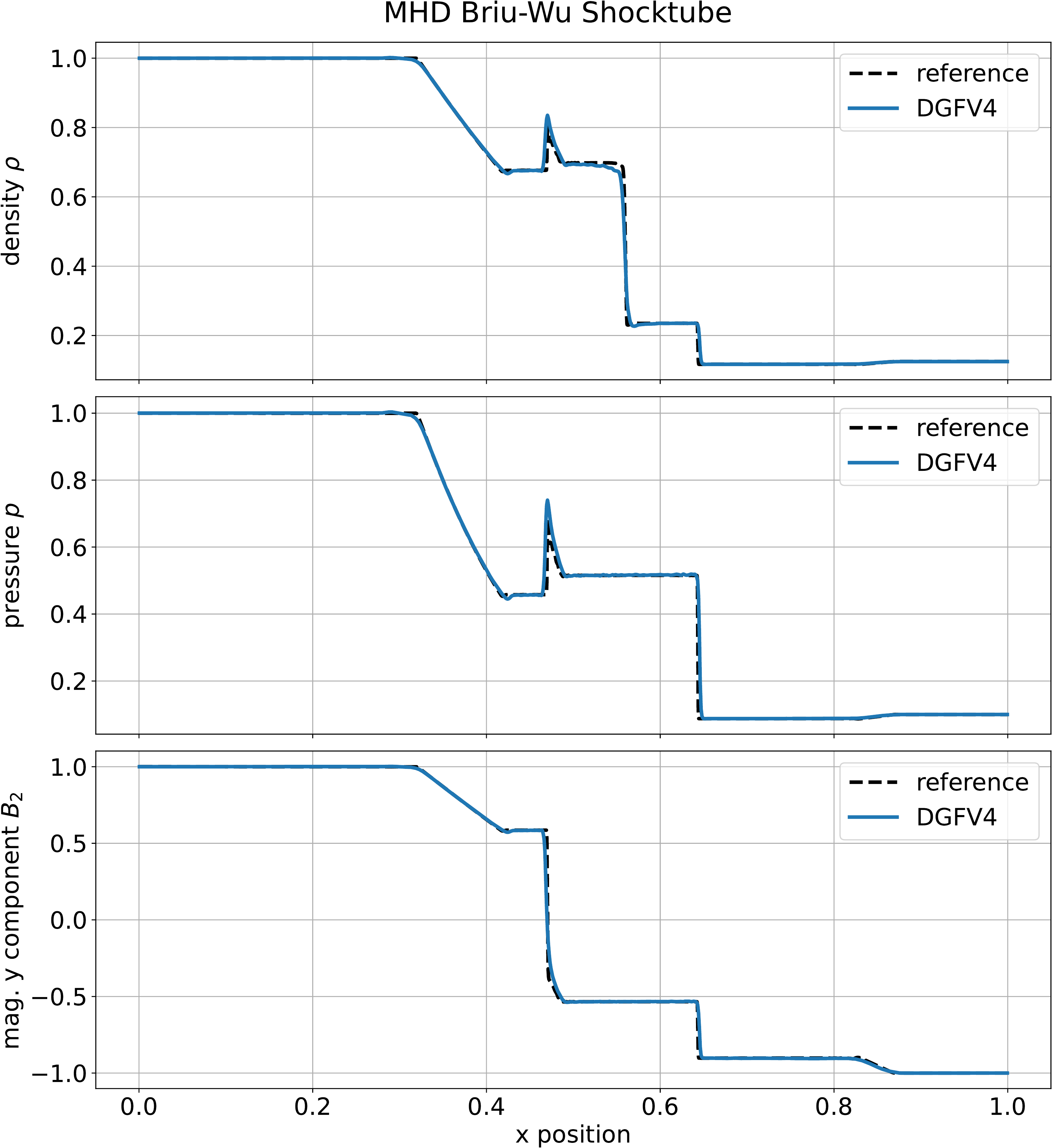}
\caption{Density, pressure and magnetic field profiles of the Briu-Wu shock tube
at final time $T = 0.1$ run by the DGFV4 scheme on an AMR grid with a minimum resolution of 32 cells and a maximum resolution of
512 cells. We compute the reference solution with the
ATHENA code using a resolution of 2048 cells.}
\label{fig:briu-wu-profiles}
\end{figure}

\Figref{fig:briu-wu-profiles} presents the density, pressure, and $y$-component
of the magnetic field of the Brio-Wu shock tube test problem at final time $T =
0.1$, using the DGFV4 scheme on an AMR grid with a minimum resolution of 32 cells and a maximum resolution of
512 cells. The reference solution is obtained using the latest version of
ATHENA (v4.2) \citep{stone2008athena} with default settings on 2048
cells. Our DGFV4 scheme captures all the MHD waves correctly, and the limiter
sharply resolves the shocks and the contact discontinuity within very few
cells. The limiter sufficiently suppresses overshoots and oscillations around
shocks. Furthermore, our scheme cooperates well with the standard AMR method
provided by FLASH (not directly visible in \Figref{fig:briu-wu-profiles}) and
properly traces the shock fronts at the highest resolution possible.

\subsubsection{MHD Shu-Osher Shock Tube}
The MHD version of the 1D Shu-Osher shock tube test \citep{shu1988efficient}
proposed by \citet{susanto2014high} becomes increasingly popular
\citep{derigs2016novel,guillet2019high} to test the scheme's ability to resolve
small-scale flow features in the presence of strong shocks in ideal
MHD regimes.

The setup follows the interaction of a supersonic shock wave with smooth density
perturbations. The computational domain is $\Omega = [-5,5]$ with outflow
boundary conditions. At $t = 0$, the shock interface is located at $x_0 = -4$.
In the region $x \leq x_0$, a smooth supersonic flow is initialised with
primitive states given by $(\rho, \vec{v}, p, \vec{B}, \Psi) = (3.5, 5.8846,
1.1198, 0, 42.0267, 1, 3.6359, 0, 0)$. In the rest of the domain $x > x_0$,
smooth stationary density perturbations are setup in primitive state as $(1 +
0.2 \sin(5\,x), 0, 0, 0, 1, 1, 1, 0,0)$. 
The flow is evolved until the final time $T = 0.7$.

\begin{figure}
\centering
\includegraphics[width=1.0\columnwidth]{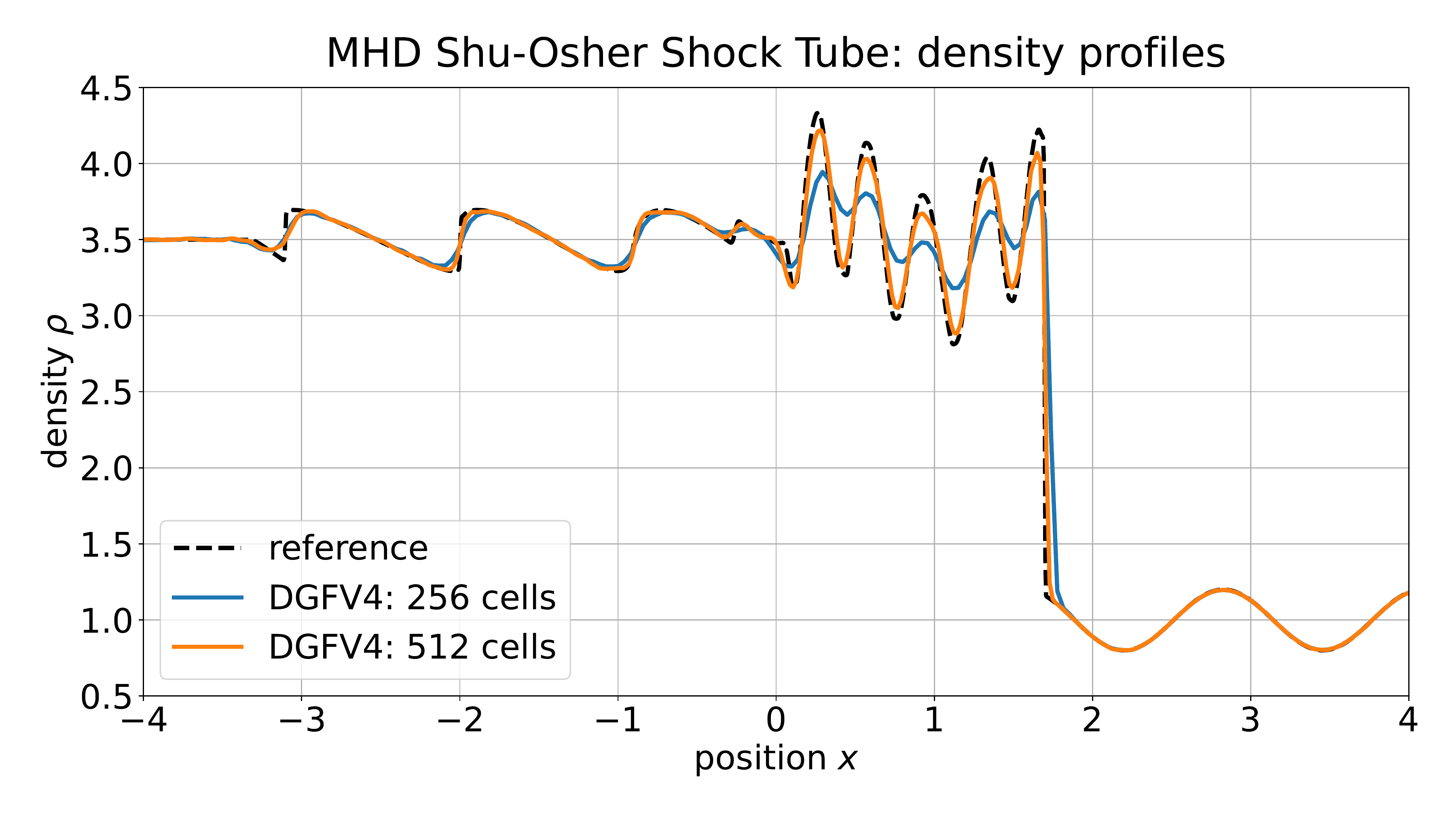}
\caption{MHD Shu-Osher shock tube test problem: The density profile is shown at
final time $T = 0.7$. The reference solution has been computed using ATHENA on 2048 grid cells.}
\label{fig:shu-osher-density}
\end{figure}

The resulting density profile at the final time $T$ is shown in
\Figref{fig:shu-osher-density}, for two maximum resolution levels: 256 cells and
512 cells.  As in the Briu-Wu shock tube, we employ AMR in order to confirm the
correct tracking of the small-scale flow features and the smooth interplay with
our MHD solver. The reference solution has been computed using ATHENA on 2048 grid cells. Our scheme properly resolves all expected flow
features for both levels of resolutions shown.

\subsubsection{Two-component Sod Shock Tube}
\begin{figure}
\centering
\includegraphics[width=1.0\columnwidth]{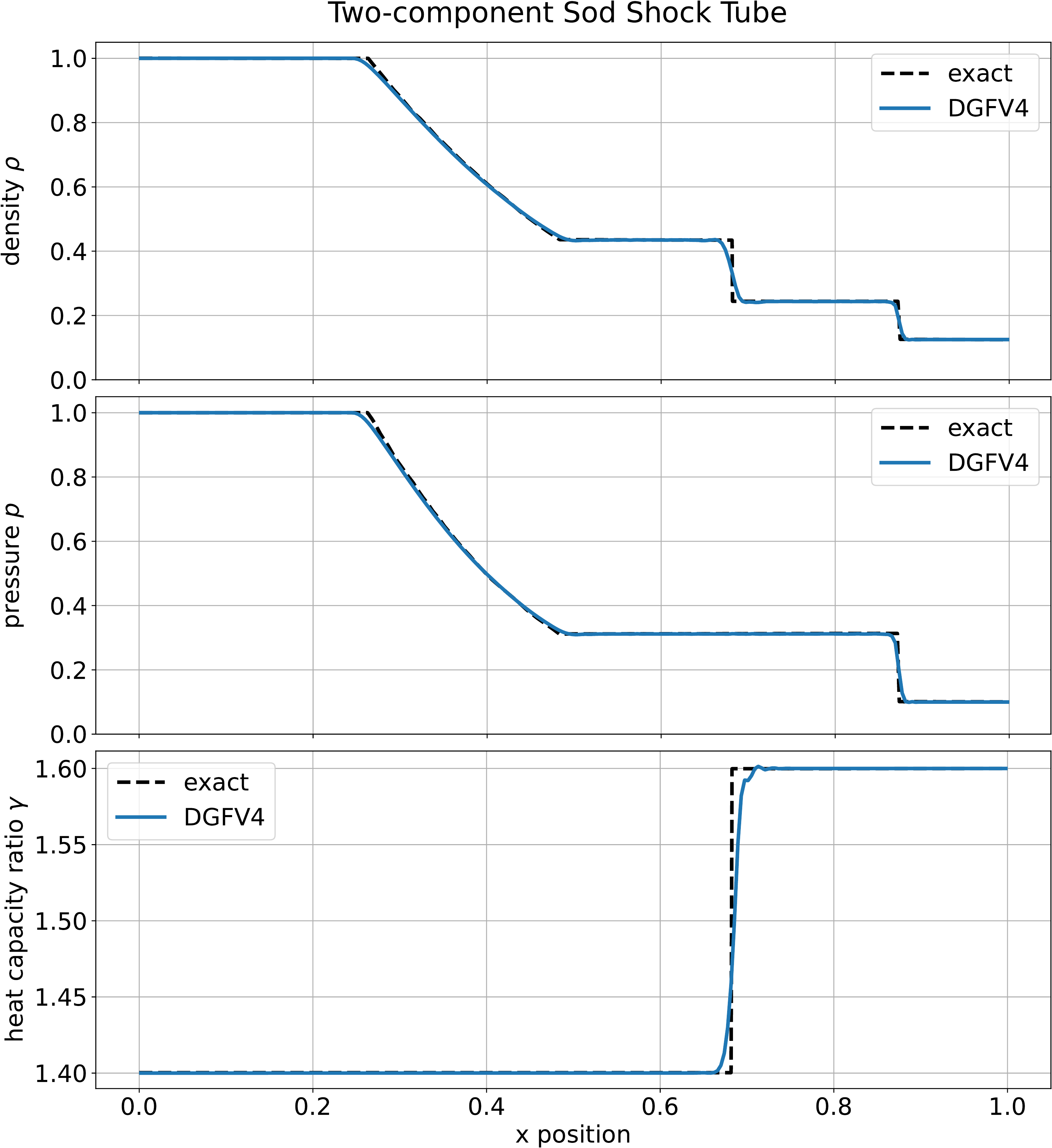}
\caption{Two-component Sod shock tube problem: The profiles of the numerical
solution with the DGFV4 scheme (on an AMR grid with a minimum resolution of 32 cells and a maximum resolution of
256 cells) together with the exact solution are
shown at final time $T = 0.2$.}
\label{fig:two-component-sod-profiles}
\end{figure}

We simulate a Sod shock tube problem extended with two species as done for
example in \cite{gouasmi2020formulation}. The computational domain is $\Omega =
[0,1]$ with outflow boundary conditions.  At $t = 0$, the shock interface is
located at $x_0 = 0.5$. In the region $x \leq x_0$, the fluid is initialised
with primitive states given by $(\rho, v, p, \sigma_1, \sigma_2) = (1, 0, 1, 1,
0)$. In the rest of the domain $x > x_0$, we setup the primitive states $(\rho,
v, p, \sigma_1, \sigma_2) = (0.125, 0, 0.1, 0, 1)$. The individual heat
capacity ratios of the two fluid components are $\gamma_1 = 1.4$ and $\gamma_2 = 1.6$
with equal heat capacities for constant volume $c^{\text{vol.}}_1 =
c^{\text{vol.}}_2 = 1$. The shock tube is evolved until the final time $T = 0.2$ run by the DGFV4 scheme on an AMR grid with a minimum resolution of
32 cells and a maximum resolution of 256 cells.
\Figref{fig:two-component-sod-profiles} shows the total density, pressure and
specific heat ratio profiles at the final time. There is an excellent
agreement with the exact solution provided by \citet{karni1994multicomponent}.

\subsubsection{Orszag–Tang Vortex}
Now we look at the 2D Orszag–Tang vortex problem \citep{orszag1979small}, a
widely-used test problem for ideal MHD. The vortex starts from a smooth initial
field configuration, and quickly forms shocks before transitioning into turbulent
flow. For this problem, our computational
domain is $\Omega = [0,1]^2$ and we use $\gamma = 5/3$. The initial density and
pressure are uniform, $\rho = 1$ and $p = 1/\gamma$. The initial fluid velocity
is $\vec{v} = (-\sin(2\pi\,y), \sin(2\pi\,x), 0)^T$, and the initial magnetic
field is $\vec{B} = (-\sin(2\pi\,y), \sin(4\pi\,x), 0)^T/\gamma$.

\begin{figure}
\centering
\includegraphics[width=1.0\columnwidth]{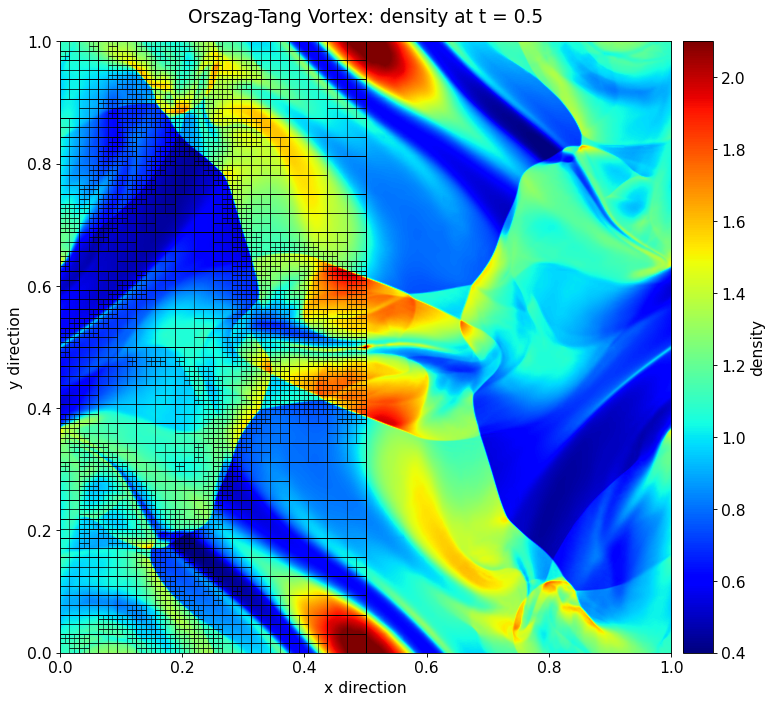}
\caption{Density profile of the Orszag-Tang vortex at the final time $T = 0.5$
as calculated with our DGFV4 scheme on an AMR mesh (shown as black lines in the
left half of the plot) with a maximum refinement level of $l = 8$, which is
equivalent to a maximum resolution of $1024^2$ cells.}
\label{fig:orszag-tang-density}
\end{figure}

The final solution at time $T = 0.5$ is shown in
\Figref{fig:orszag-tang-density}.  We employ an AMR grid with resolution levels
going from $64^2$ cells to $1024^2$ cells, which is highlighted as
black lines on the left half of the density plot.  We recognise the well-known
density distribution of the Orszag–Tang vortex, which is commonly presented in
MHD code papers. Note that we obtain both sharp shocks and smooth, noise-free
flow with resolved features between the shocks.

\begin{figure}
\centering
\includegraphics[width=1.0\columnwidth]{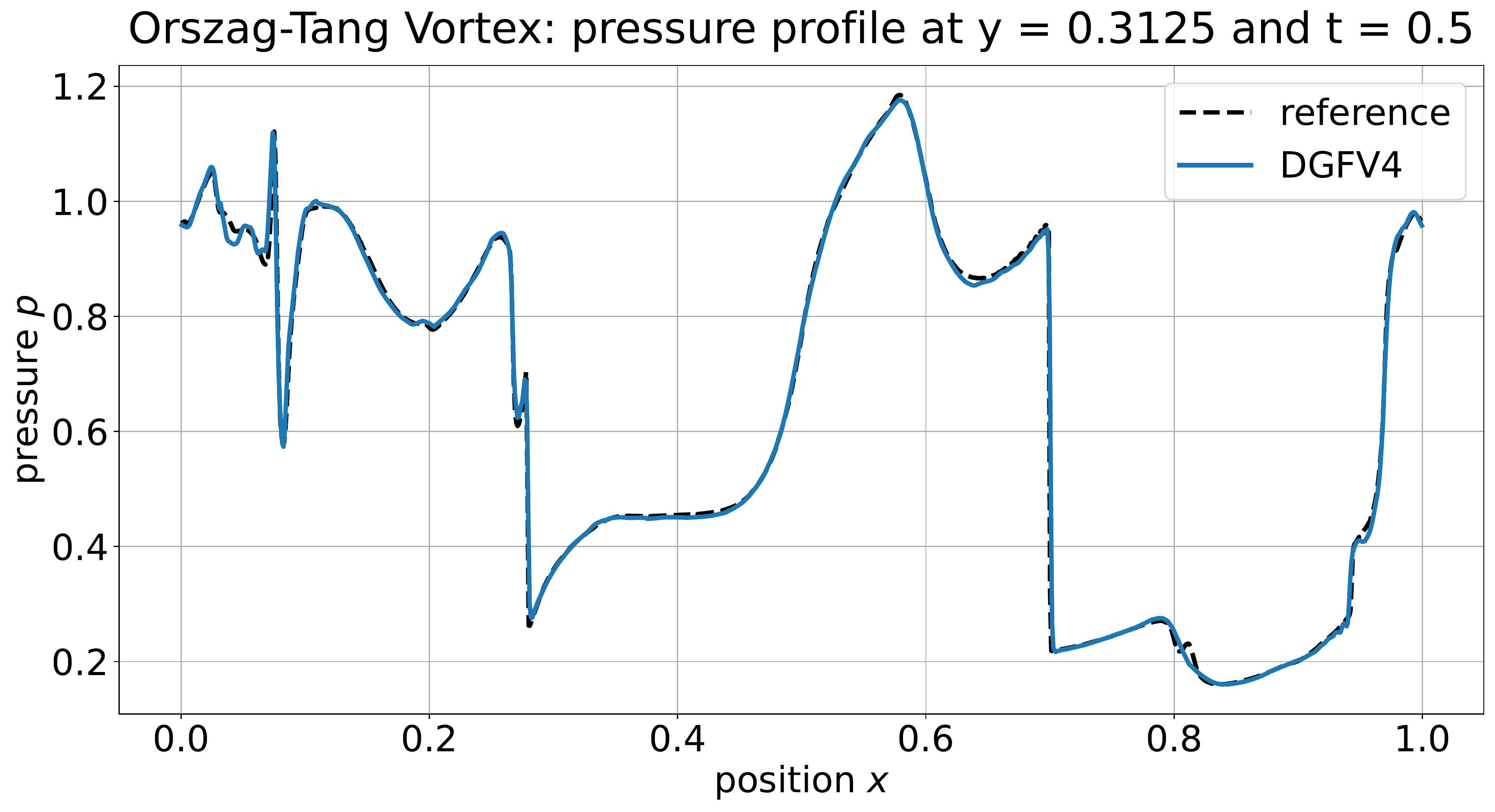}
\caption{Numerical solution of the pressure profiles at $y = 0.3125$ and time
$T = 0.5$ are shown for the DGFV4 solver with AMR and a maximum resolution of
$1024^2$ cells.  The reference solution has been computed with the ATHENA code at a
uniform resolution of $2048^2$ cells.}
\label{fig:orszag-tang-pressure-profile}
\end{figure}

In order to corroborate the correct positioning of the shock waves, we computed
a reference solution on a finer grid of $2048^2$ cells with the ATHENA code and
overlay 1D cuts of both pressure solutions in
\Figref{fig:orszag-tang-pressure-profile}. The resulting profile of our scheme
matches the reference very well and is very sharp and without spurious
oscillations or overshoots. 

\begin{figure}
\centering
\includegraphics[width=1.0\columnwidth]{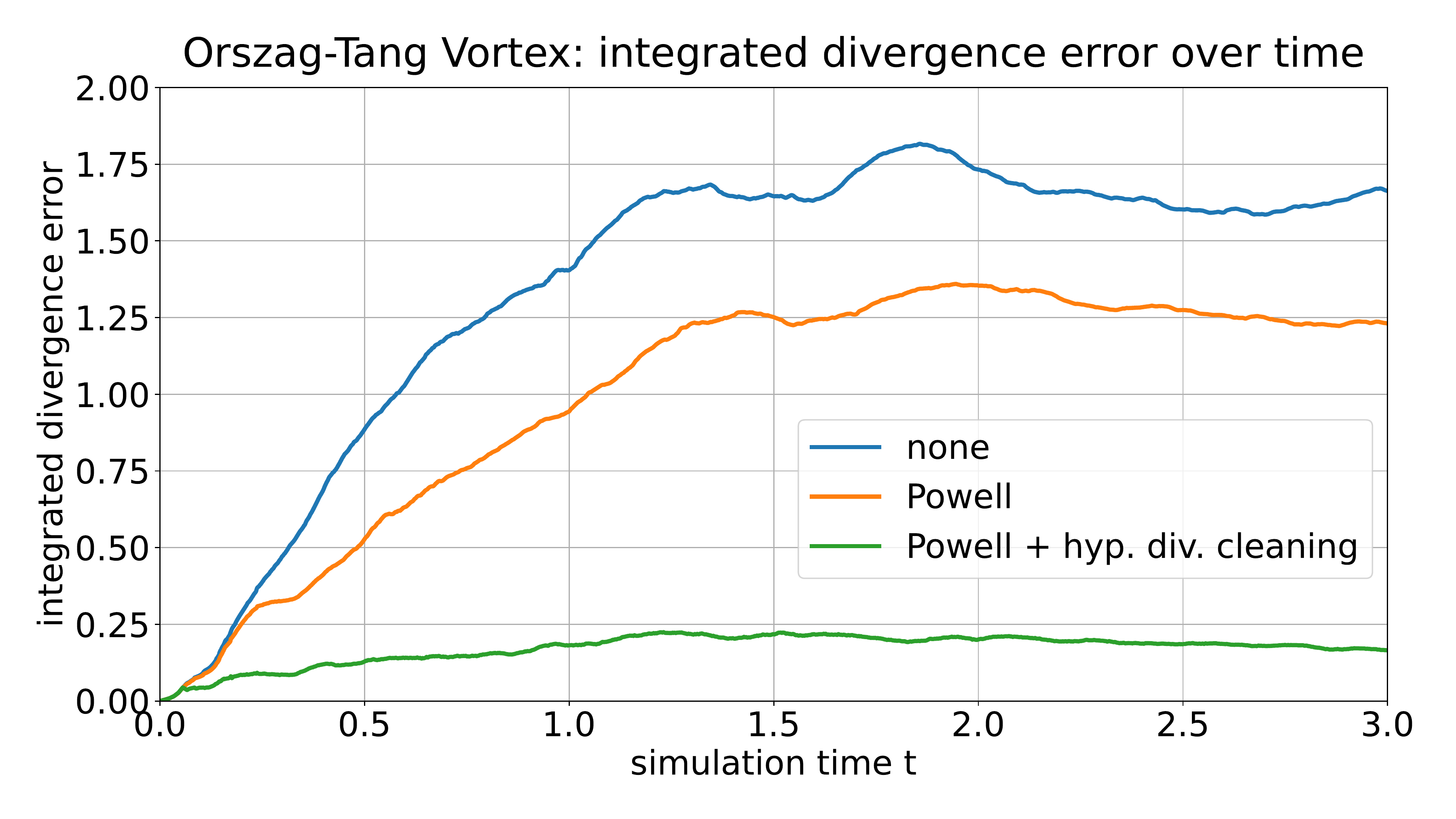}
\caption{Domain-integrated divergence error $\int_{\Omega} |\nabla\cdot\vec{B}|
d\Omega$ as a function of the simulation time for the Orszag-Tang vortex setup. Three runs with
the DGFV4 are shown where the Powell scheme and hyperbolic divergence cleaning
are either on or off. The simulation without any divergence correction method
activated is labelled  as "none".}
\label{fig:orszag-tang-divergence-over-time}
\end{figure}

During our numerical experiments we found that without proper divergence
cleaning, distinctive grid artefacts appear, which considerably pollute the
solution.  Analog to \Secref{sec:current-sheet} we compared the time evolution
of the divergence errors with and without divergence control mechanisms. The
results shown in \Figref{fig:orszag-tang-divergence-over-time} clearly
demonstrate that the hyperbolic divergence cleaning method is effective in
confining the divergence errors ensuring a clean, unpolluted MHD flow.

\subsubsection{Magnetic Rotor}
We now investigate the 2D MHD rotor problem introduced by
\citet{balsara1999staggered}. In this setup, a dense disc of fluid rotates
within a static fluid background, with a gradually declining velocity layer
between the disk edge and the ambient fluid.  An initially uniform magnetic
field is present, winding up with the disc rotation and containing the dense
rotating region through magnetic field tension. The computational domain is set
to $\Omega = [0, 1]^2$. Initial pressure and magnetic fields are uniform in the
whole domain, with $p = 1$ and $\vec{B} = (5/4\pi, 0, 0)^T$. The central,
rotating disc is defined by $r < r_0$ where $r^2 = (x - 0.5)^2 + (y - 0.5)^2$,
and $r_0 = 0.1$. Inside the disc, $\rho = 10$, and the disc rotates rigidly
with $\vec{v} = (0.5-y, x-0.5)v_0/r_0$ with $v_0 = 2$.  The background fluid
has a density of $\rho = 1$ and is at rest: $\vec{v} = \vec{0}$. In the annulus
$r_0 \leq r \leq r_1 = 0.115$, the transition region linearly interpolates
between the disc and the background, with $\vec{v} = (0.5-y, x-0.5)v_0 f /r_0$
and $\rho = 1+9f$, where $f = (r_1 -r)/(r_1 -r_0)$ is the transition function.
The simulation runs until the final time $T = 0.15$. We use outflow boundary
conditions.

\begin{figure}
\centering
\includegraphics[width=1.0\columnwidth]{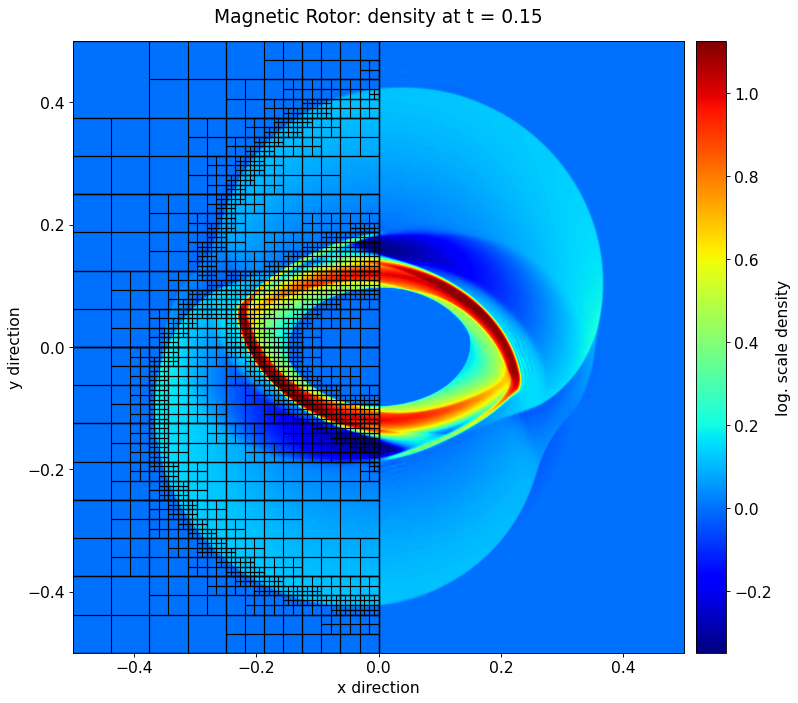}
\caption{Density contours in logarithmic scale of the magnetic rotor at final
time $T = 0.15$ calculated with DGFV4 on an dynamic AMR mesh (shown as black
lines in the left half of the plane) with a maximum refinement level of $l = 8$
which is equivalent to a maximum resolution of $1024^2$ cells.}
\label{fig:magnetic-rotor-density}
\end{figure}

We present the density contours at $T = 0.15$ in
\Figref{fig:magnetic-rotor-density}. The black lines in the left half the plot
highlight the AMR grid bounded between refinement levels of $64^2$ cells to
$1024^2$ cells. The contours of the evolved disc show a sharp and noise-free
picture of the circular rotation pattern, which is in general a challenge for
MHD codes.

\begin{figure}
\centering
\includegraphics[width=1.0\columnwidth]{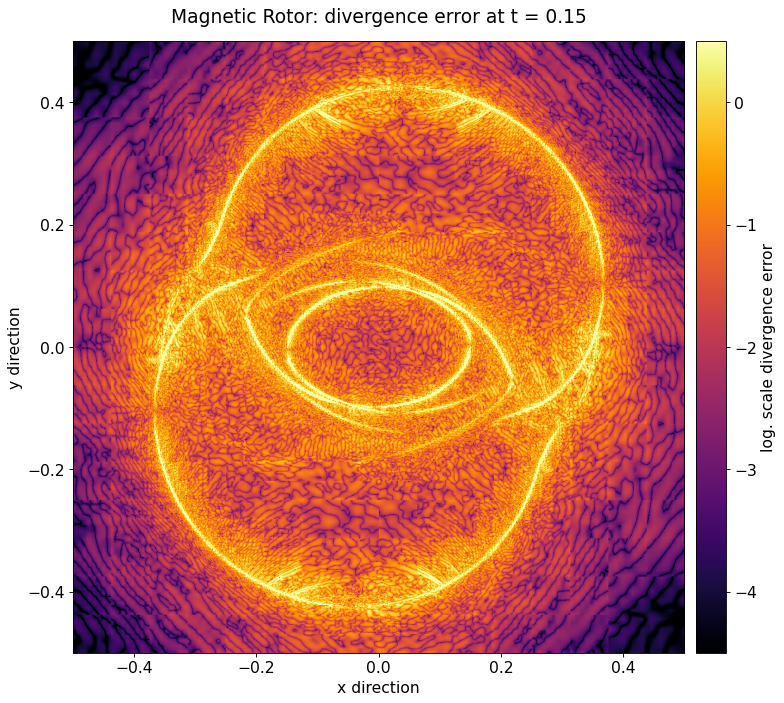}
\caption{Contours of the local divergence error $|\nabla\cdot\vec{B}|$ in
logarithmic scale of the magnetic rotor run by the DGFV4 scheme at final time $T = 0.15$.}
\label{fig:magnetic-rotor-divergence}
\end{figure}

In \Figref{fig:magnetic-rotor-divergence}, we plot the local magnetic field
divergence error $|\nabla\cdot\vec{B}|$ of the DGFV4 solver at the final simulation time. The error is mostly concentrated in regions around shocks and radially propagates away in
all directions (circular ripples) due to the hyperbolic divergence advection
mechanism detailed in \Secref{sec:hyperbolic-div-cleaning}. Preferably, the
divergence error gets advected out of the domain but, as clearly visible in the
divergence plot, the errors can accumulate in stagnant regions of the domain,
which justifies the importance of the damping source term
\eqref{eq:GLM-source-damp} as an additional mechanism to dispose of any accrued
magnetic field divergence errors.

\subsubsection{MHD Blast}
In order to test the shock-capturing performance of our code for a very strong
ideal MHD shock problem in 3D, we utilise the 3D blast wave setup of
\citet{balsara2009efficient}. The computational domain is set to $\Omega = [0,
1]^3$ with outflow boundary conditions. The background fluid is initially at
rest with respect to the grid, where $\vec{v} = \vec{0}$, $\rho = 1$, and with
a uniform magnetic field $\vec{B} = (100\sqrt{3},100\sqrt{3},0)$. The ambient
pressure is set to $p =  10^{-1}$, and within a central sphere of radius $r_0 =
0.1$, we set $p = 10^{3}$ to initialise the blast. This creates an extreme
initial shock strength with a pressure ratio of $10^4$ in a strongly
magnetised background with a plasma-$\beta$ of $\approx10^{-5}$ . We take
$\gamma = 1.4$, and run the simulation until the final simulation time $T =
0.01$. The AMR grid is bounded between refinement levels of $64^3$ cells to
$256^3$ cells.

\begin{figure}
\centering
\includegraphics[width=1.0\columnwidth]{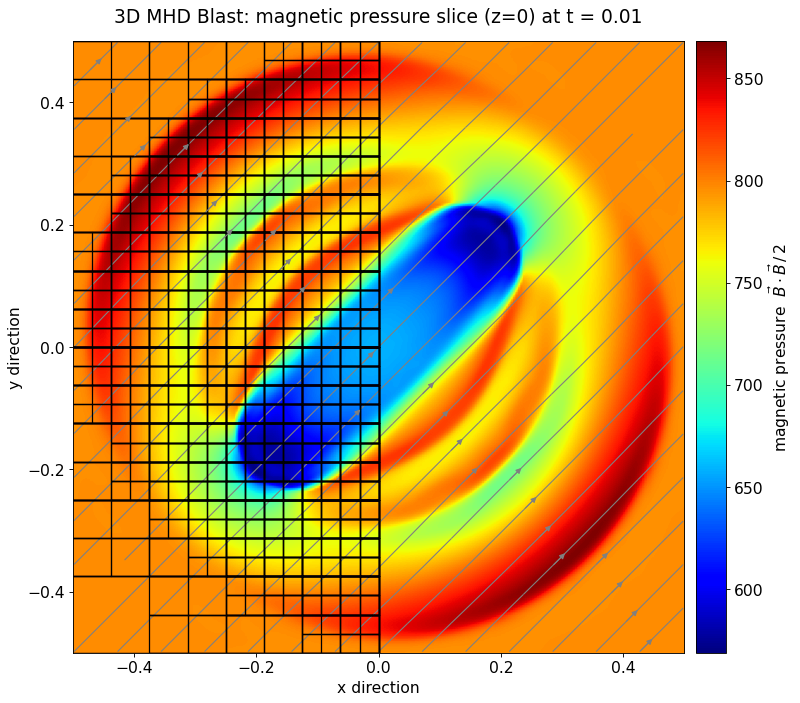}
\caption{Slice of the magnetic pressure at $z = 0$ for the MHD blast wave test
solved with the DGFV4 scheme. The black lines on the left side highlight part
of the AMR mesh and the grey stream lines indicate the field spanned by the
$x$- and $y$-component of the magnetic field.}
\label{fig:mhd-blast-magnetic-pressure}
\end{figure}

In \Figref{fig:mhd-blast-magnetic-pressure}, we show a slice at $z = 0$ of the
magnetic pressure at the final time. The black lines in the left half of the
plot highlight the AMR grid, while the light grey arrows trace the magnetic
field lines of the solution. The originally spherical blast bubble gets
stretched along the magnet background field over the course of the simulation
as is described by \citet{balsara2009efficient}.
Our scheme is able to maintain positivity of the pressure and density in the
whole domain and correctly captures the very strong discontinuities, while
resolving the complex structures within the blast shell. Note that no
oscillations are visible around discontinuities. This test shows the robustness
and shock-capturing performance of our blending scheme for three-dimensional
problems involving very strong magnetised shocks.

\subsection{Coupling to Gravity}
\label{sec:coupling-gravity}
The FLASH framework provides specialised units for the employment of static
gravitational sources emanating from point masses distributed at arbitrary
points within the computational domain or from external background fields.
Furthermore, FLASH unites a variety of solvers for the Poisson equations of
gravity for mass distributions (self-gravity) based on multi-pole expansions
\citep{muller1995simulating}, tree-based Barnes-Hut algorithms (BHTree)
\citep{barnes1986hierarchical, wunsch2018tree}, multi-grid methods
\citep{ricker2008direct} or hybridized schemes (Pfft)
\citep{daley2012optimization}.

At first, we show an interesting MHD simulation using the static point mass
module in FLASH and in the second part we verify the proper coupling of our
code to the gravity solvers in FLASH for the case of the tree solver
implementation by \citet{wunsch2018tree}.

\subsubsection{Differentially Rotating MHD Torus}
\label{sec:mhd-torus}
Here, we present the results of a three-dimensional global MHD simulation of
magneto-rotationally instabilities (MRI) growing in a differentially rotating
torus initially threaded by a toroidal magnetic field \citep{okada1989model}.
After several rotation periods the inner region of the torus becomes turbulent
due to the growth of the MRI fuelling efficient angular momentum transport
processes
\citep{velikhov1959stability,chandrasekhar1960stability,balbus1990powerful}.
Magnetic loops emerge by the buoyant rise of magnetic flux sheets from the
interior of the torus \citep{machida1999three}, also known as Parker instability
\citep{parker1966dynamical}. Owing to this angular momentum
redistribution, the torus becomes flattened to a disc where interstellar
material gradually accretes to the massive source of gravity at the centre.
Consequently, this process leads to various interesting phenomena such as X-ray
flares and jet formation.

\begin{figure*}
\centering
\includegraphics[width=1.8\columnwidth]{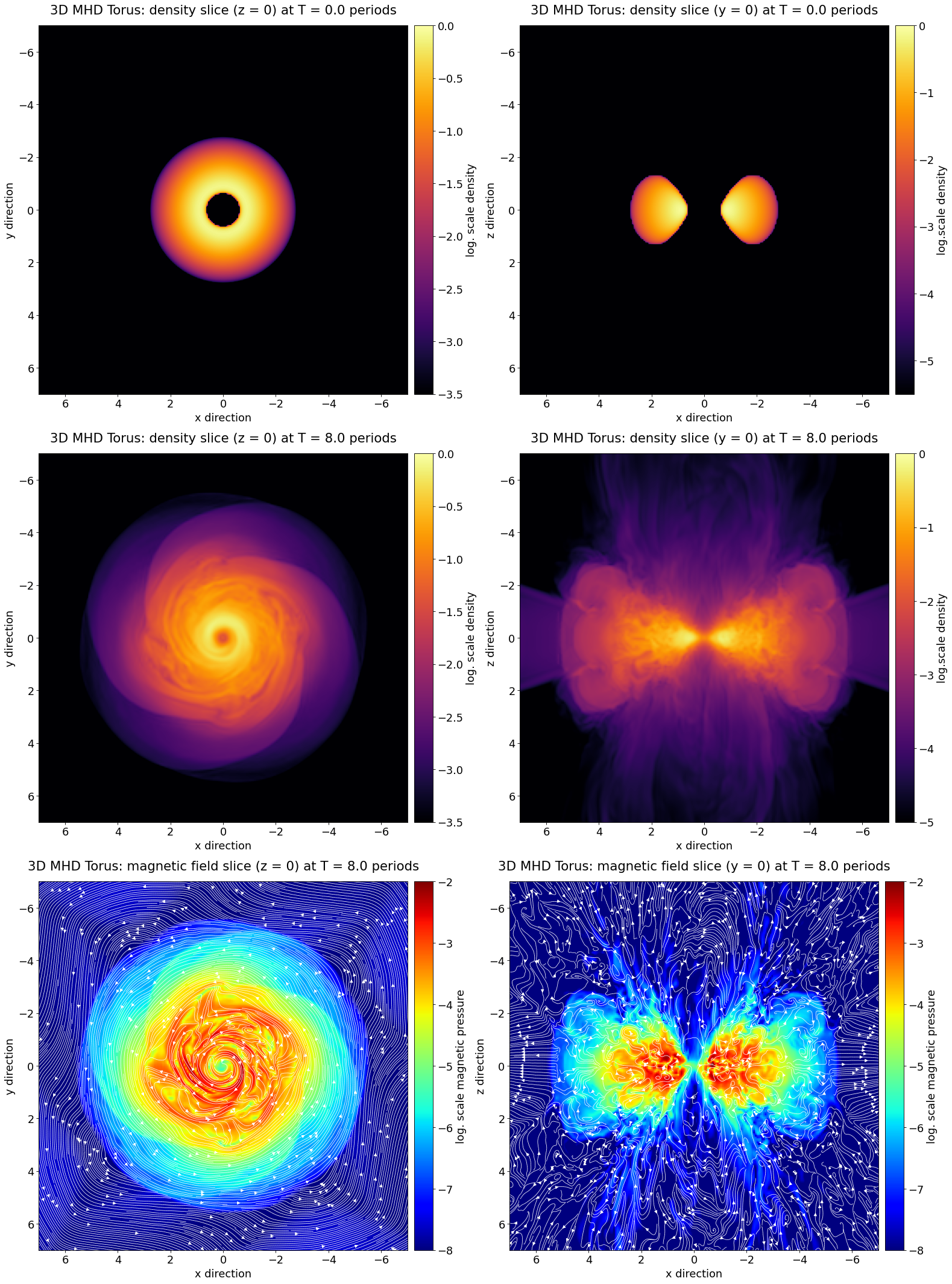}
\caption{MHD torus at the initial time (top row) and after 8 periods: density
(middle row) and magnetic field (bottom row) slices.}
\label{fig:torus-slices}
\end{figure*}

Such setups are usually modelled with the ideal MHD equations in cylindrical
coordinates $(r,\phi,z)$, but in this work we resort to the Cartesian
coordinate system $(x,y,z)$. Following \citet{machida1999three}, we adopt an
equilibrium model of the magnetised torus in order to initialise the setup. The
scales of the model are completely determined by the following parameters:
gravitational constant $G$, mass of the central object $M$, initial angular
momentum of the torus $L$, median radius of the torus $r_0$, characteristic
density of the torus $\rho_0$, magnetic field strength of the initial toroidal
magnetic field $B_{\phi}$ and the initial ratio of gas pressure to magnetic
pressure $\beta_0$.  For convenience we set all aforementioned parameters to
unity: $G = M = L = r_0 = \rho_0 = B_{\phi} = \beta_0 = 1$. We assume the
polytropic equation of state $p = K\,\rho^{\gamma}$ with constant $K = 0.05$
and $\gamma = 5/3$.  The inviscid, non-resistive torus is embedded in
non-rotating, non-self-gravitating ambient gas ($\rho = 10^{-5}$) and evolves
adiabatically. Radiative cooling effects are neglected. For the static
gravitational field caused by the central mass, we use the Newtonian potential
provided by the {\it PointMass} module.  The Alfvén
wave speed is calculated according to
\citet{okada1989model} as
\begin{equation}
    c_{\text{Alfvén}}^2 = \frac{2\,K}{\beta_0}\big(\rho\,r^2\big)^{\gamma-1} \;\;\text{with}\; r^2 = x^2 + y^2,
\end{equation}
and the sound speed $c_{\text{sound}}$ is defined by \eqref{eq:wave-speeds}.
We transform the equation of motion \citep{machida1999three} to the potential
form
\begin{equation}
\label{eq:torus-potential}
    \psi_{\text{torus}} = \text{const.} = -\frac{G\,M}{r} + \frac{L^2}{2\,r^2} + \frac{c_{\text{sound}}^2}{\gamma-1} + \frac{\gamma\,c_{\text{Alfvén}}^2}{2\,(\gamma-1)}\;\Bigg|_{r\,=\,r_0}
\end{equation}
and obtain an expression for the density distribution of the initial torus
\begin{equation}
\label{eq:torus-density}
    \rho_{\text{torus}} = \left[\frac{\max\big(0,\psi_{\text{torus}} + G\,M/R-L^2/(2\,r^2)\big)}{K\,\gamma/(\gamma-1)\big(1 + r^{2\,(\gamma-1)}/\beta_0\big)}\right]^{1/(\gamma-1)}.
\end{equation}
The torus rotates differentially according to the Keplerian velocity $v_{\phi}
= \sqrt{G\,M/r}$. The computational domain is set to $\Omega = [-7,7]^3$ with outflow
boundary conditions at all faces. The AMR grid is bounded between refinement levels of $64^3$ cells to $256^3$ cells. The first row in \Figref{fig:torus-slices}
shows what the initial torus looks like in our setup.

After several periods, regions of dominating magnetic pressure $(\beta < 1)$
emerge, where  magnetic flux buoyantly escapes from the disk leading to violent
flaring activities. The loop-like structures, similar to those in the solar
corona, are also visible in our simulation after eight rotation periods as shown
in \Figref{fig:torus-slices} in the second and third row. This setup shows
that the DGFV4 solver is capable of simulating complex MHD flow configurations under the influence of a static gravitational potential, giving rise
to magnetically driven structures over several spatial scales. 

We further compare our implementation
with another established MHD solver readily available in FLASH and perform simulations with different resolutions. The results can be found in appendix~\ref{sec:comp-study-torus}.

\subsubsection{Evrard Test}
The Evrard test described by \citet{evrard1988beyond} investigates the
gravitational collapse and subsequent re-bounce of an adiabatic, initially cold
sphere. It is generally used to verify energy conservation when hydrodynamics
schemes are coupled to gravity \citep{springel2001gadget,wetzstein2009vine}.
The initial conditions consist of a gaseous sphere of mass $M$, radius $R$, and
density profile
\begin{equation*}
\rho(r) = \frac{M}{2\pi\,R^2\,r}.
\end{equation*}
The uniform temperature is initialised so that the internal
energy per unit mass is
\begin{equation*}
E_{\text{int}} = 0.05\,\frac{G\,M}{R},
\end{equation*}
where $G$ is the gravitational constant. The standard values of the above
parameters, used also in this work, are $M = R = G = 1$. The Barnes-Hut MAC is
set to $\theta_{\text{lim}} = 0.5$ and the AMR refinement levels range from
$64^3$ cells to $256^3$ cells. We computed the reference solution on a uniform
grid of $256^3$ cells with the PPM solver for
hydrodynamics \citep{colella1984piecewise} and the same tree solver for gravity with equal settings as
presented in \citet{wunsch2018tree}.

\begin{figure}
\centering
\includegraphics[width=1.0\columnwidth]{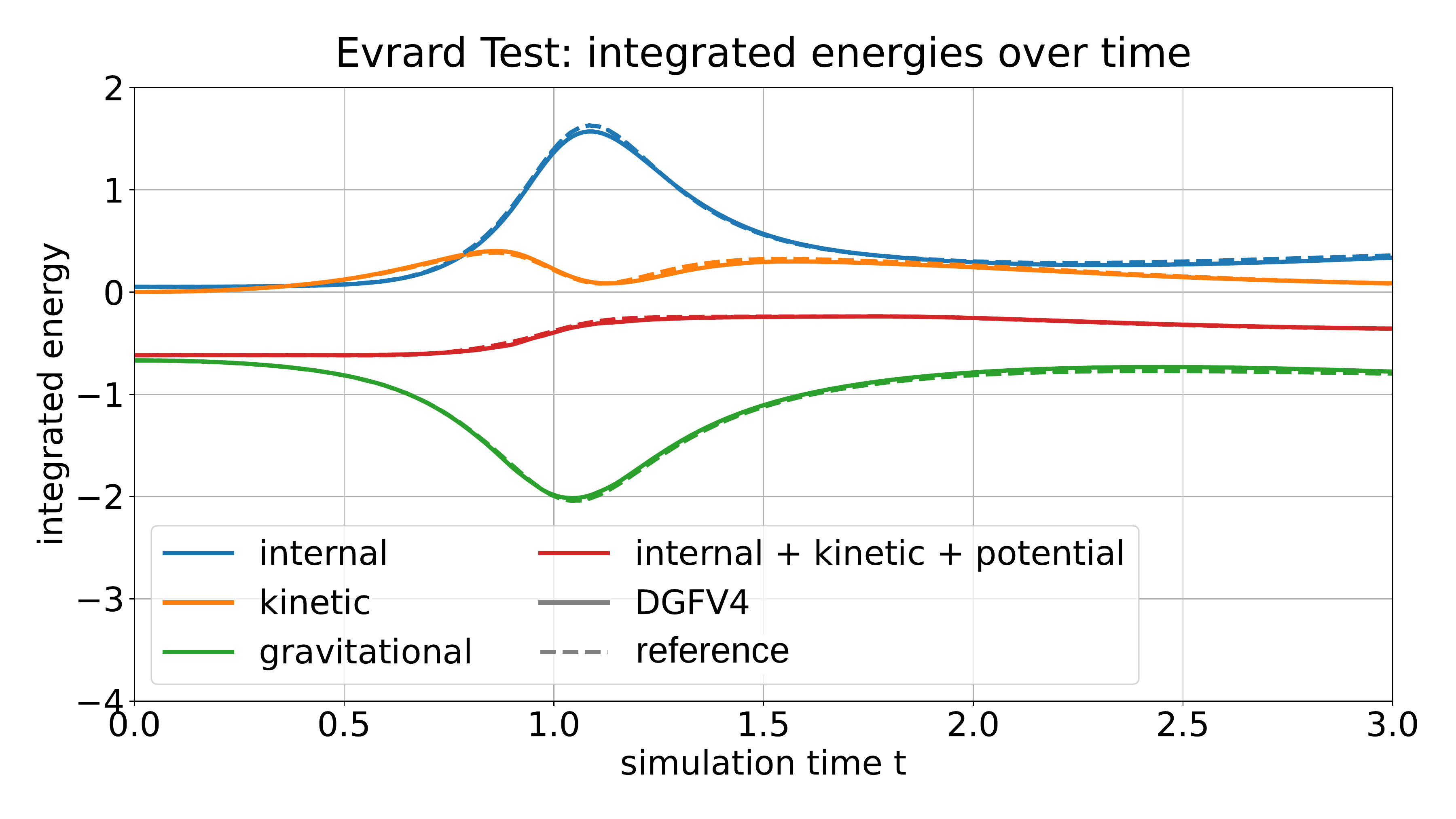}
\caption{Domain-integrated energies over simulation time for the Evrard test run by
our DGFV4 solver. The dashed lines show the reference solution computed with
PPM on a uniform grid.}
\label{fig:evrard-energies-over-time}
\end{figure}

\Figref{fig:evrard-energies-over-time} presents the time evolution of the
domain-integrated gravitational, kinetic, internal, and the sum of the three energies.
Our results match the reference very well which confirms the correct coupling
to the gravity solver. Since the interface to the gravitational source terms in
FLASH is completely generic our findings transfer to any other gravity solver
module available in FLASH. Note that the reason for a deviation from energy
conservation has been discussed extensively in \citet{wunsch2018tree}. The
source of the inaccuracy is the finite grid resolution, which leads to an error
at the point in time where the sphere is most compressed (least resolved) and
continues to bounce back.

%%%%%%%%%%%%%%%%%%%%%%%%%
\subsection{Coupling to TreeRay}
\label{sec:coupling-treeray}
Turbulent, multi-phase structures within the interstellar medium (ISM) are
shaped by the complex and non-linear interplay between gravity, magnetic
fields, heating and cooling, and the radiation and momentum input from stars
\citep[e.g.][]{agertz2013toward,walch2015silcc,kim2017modeling}.

TreeRay \citep{wunsch2021tree} is a new radiation transport method combining an
octree \citep{wunsch2018tree} with reverse ray tracing. The method is currently
implemented in FLASH. In short, sources of radiation and radiation absorbing
gas are mapped onto the tree encoded as emission and absorption coefficients.
The tree is traversed for each grid cell and tree nodes are mapped onto rays
going in all directions. Finally, a one-dimensional radiation transport
equation is solved along each ray. Several physical processes are integrated
into the TreeRay code by providing prescriptions for the absorption and the emission
coefficients.

In this work, we employ a simple On-the-Spot approximation with only one source
(a massive star) emitting a constant number of extreme ultraviolet (EUV)
photons per unit time.  The Spitzer sphere \citep{spitzer1978physical} is a
simple model regarding the interaction of ionizing radiation with absorbing
gases. In this model, the EUV radiation from a young, massive star ionizes and
heats the surrounding medium, creating a so-called HII region, i.e. an
over-pressured, expanding bubble of photo-ionized gas bounded by a sharp
ionisation front \citep{whitworth1979erosion,deharveng2008triggered}.  The
expanding ionisation front drives a shock into the surrounding neutral, cold
gas, sweeping it up into a dense, warm shell.

All radiation-hydrodynamic codes use this spherical expansion of an
HII region as a standard test problem \citep{bisbas2015starbench}. The
problem is insofar challenging, since it involves a complicated combination of
fluid dynamics, radiative transfer, micro-physical heating, cooling, ionisation
and recombination.

After we verify our simulation of the spherical expansion of an
HII region with the Starbench test (\Secref{sec:starbench}), we extend the setup in
order to simulate the expansion of an HII region into a fractal cloud
(\Secref{sec:fractal}).

\subsubsection{Starbench Test}
\label{sec:starbench}
\citet{bisbas2015starbench} introduced a standard benchmark test in 3D for
early-time ($t \leq 0.05$ Myr) and late-time ($t > 0.05$ Myr) expansion phases
of the process. In this work we test our code with the 3D late expansion phase
simulation which is coined the Starbench test.

An analytic approximation for the time evolution of the ionisation front (IF)
is given by \citet{spitzer1978physical}
\begin{equation}
\label{eq:spitzer}
    R_{\text{Spitzer}}(t) = R_{\text{S}}\left(1 + \frac{7\,c_{\text{ionized}}\,t}{4\,R_{\text{S}}}\right)^{4/7}
\end{equation}
where
\begin{equation}
    R_{\text{S}} = \left(1 + \frac{3\,\dot{N}\,m_p^2}{4\pi\,\alpha_B\,\rho_0^2}\right)^{1/3}
\end{equation}
is the Strömgren radius \citep{stroemgren1939}, $c_{\text{ionized}} =
12.85\,\text{km}\,\text{s}^{-1}$ is the sound speed in the ionised gas inside
the isothermal bubble of $T_{\text{ionised}} = 10^4\,\text{K}$.  $\dot{N} =
10^{49}\,\text{s}^{-1}$ is the rate at which the source at the centre emits
hydrogen ionising photons ($E_{\nu} > 13.6 \text{eV}$), $m_p =
1.67\cdot10^{24}$ g is the proton mass, and $\alpha_B = 2.7 \cdot 10^{-13}
\text{cm}^3 \text{s}^{-1}$ is the recombination coefficient. The density of the
surrounding neutral cloud of only hydrogen gas ($\gamma = 5/3$) is taken to be
$\rho_0 = 5.21\cdot10^{-21}\text{g}\,\text{cm}^{-3}$ and has a temperature of
$T_{\text{neutral}} = 10^3\,\text{K}$. The corresponding sound speed is then
$c_{\text{neutral}} = 2.87\,\text{km}\,\text{s}^{-1}$.  If, during the
simulation, the temperature in the neutral gas exceeds $T_{\text{shell}} = 300
\,\text{K}$, it is instantaneously cooled to $T_{\text{shell}}$. Consequently,
the shell of shock-compressed neutral gas immediately ahead of the expanding IF
is effectively isothermal. The cooling aims to limit the thickness of the shell
and make it resolvable for the numerical fluid solver. The given parameters
result in a Strömgren radius of $R_{\text{S}} = 0.3141\,\text{pc}$.  Note, that
this setup contains two fluid species, namely the neutral and ionised medium, and
that all aforementioned physical and chemical processes are implemented within
TreeRay which are completely opaque to the fluid solver.

Since the Spitzer solution \eqref{eq:spitzer} is only valid for the very early
expansion phase, \citet{bisbas2015starbench} proposed a heuristic equation
(equation (28) in \citet{bisbas2015starbench}) giving a good approximation of
the position of the ionisation front at all times
\begin{equation}
\label{eq:starbench}
    R_{\text{Starbench}} = R_{\text{II}} + \big(1 - 0.733 \, \exp(-t/\text{Myr})\big)\,(R_{\text{I}} - R_{\text{II}})
\end{equation}
where $R_\text{I}$ and $R_\text{II}$ are solutions to the ordinary differential
equations (8) and (12) in \citet{bisbas2015starbench}. For brevity we omit
the explicit definition of these and refer to the paper.

We run our simulation to final time $T = 1.5\,\text{Myr}$ and set the domain to
$\Omega = [0,3\,\text{pc}]^3$ which only represents one octant of the whole
setup. Since the model is spherically symmetric around the emitting radiation
source situated at the $\vec{x} = (0,0,0)^T$, we can speedup the simulation
considerably. The resolution ranges from $32^3$ cells to
$128^3$ cells. The boundaries of the domain touching the coordinate origin are
reflecting walls while the rest are set to outflow. The ionised bubble is
expected to occupy most of the domain by the end of the simulation.

\begin{figure}
\centering
\includegraphics[width=1.0\columnwidth]{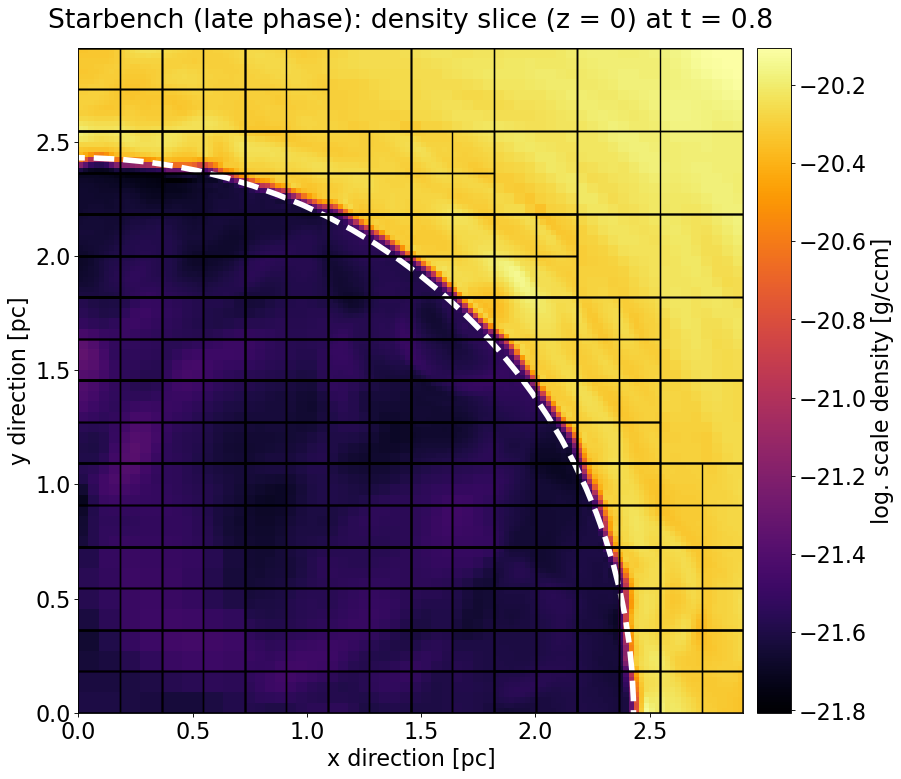}
\caption{Density slice of the Starbench test setup at simulation time $t = 0.8$
Myr computed with our DGFV4 solver and a maximum resolution of $256^3$ cells.
The dashed circle highlights the Starbench solution given by \eqref{eq:starbench}.
The black lines depict the AMR grid.}
\label{fig:starbench-density-slice}
\end{figure}

\Figref{fig:starbench-density-slice} is a snapshot of the expanding bubble at
time $t = 0.8\,\text{Myr}$ showing a density slice at constant $z = 0$. The dark
violet region of lower density is completely filled with hot gas, photo-ionised
by the radiation emitting point source sitting in the lower left corner. The
dashed white circle demarks the ionisation front separating the inner, ionised
medium from the surrounding, neutral medium of higher density. The dashed circle
is computed with the Starbench solution \eqref{eq:starbench}
confirming excellent agreement between theoretical and numerical results.

\begin{figure}
\centering
\includegraphics[width=1.0\columnwidth]{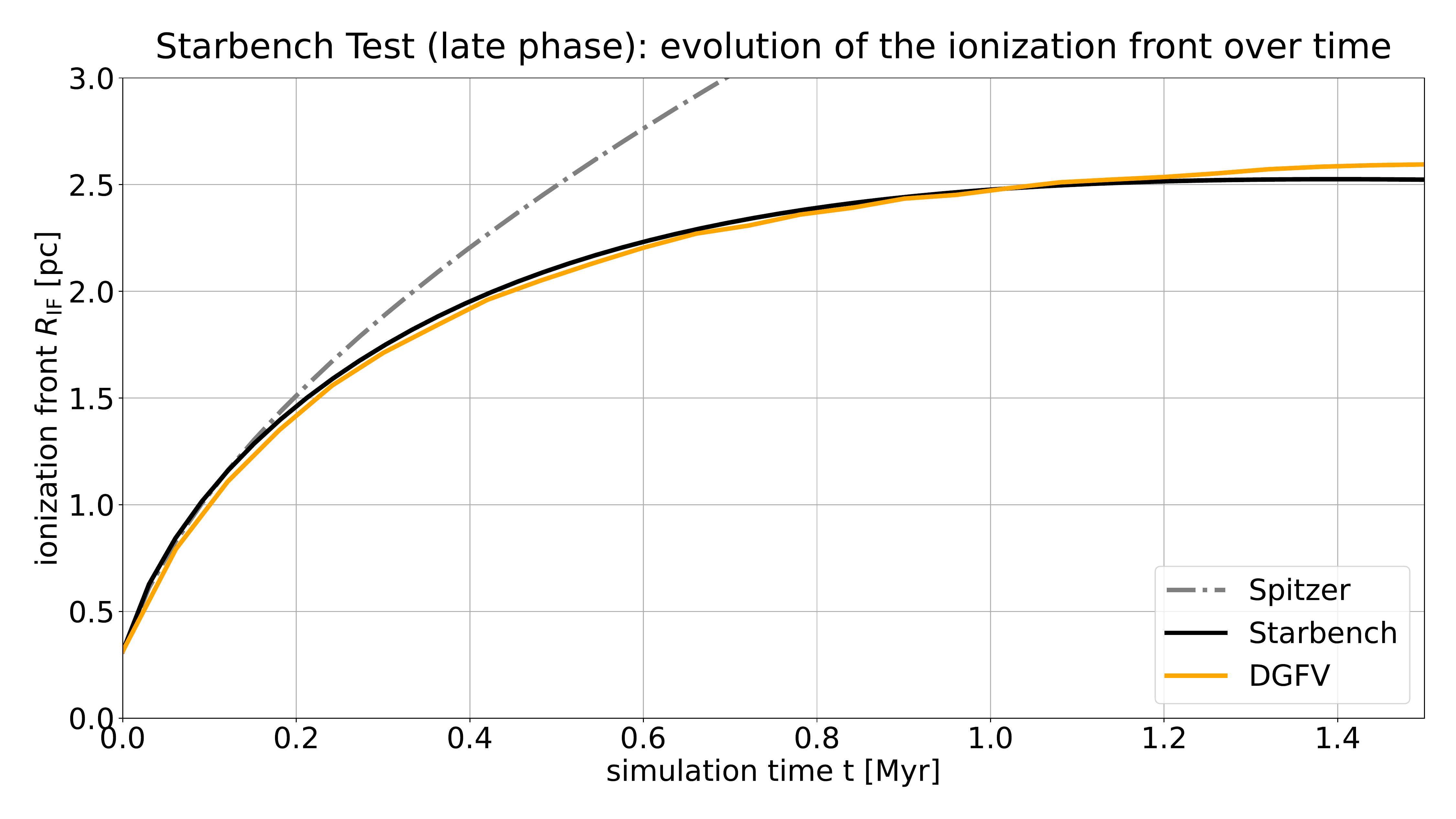}
\caption{Ionisation fronts over time computed with our DGFV4 solver. For reference the
Spitzer solution \eqref{eq:spitzer} and the Starbench solution \eqref{eq:starbench} are
plotted.}
\label{fig:starbench-if-over-time}
\end{figure}

Additionally, we plotted the position of the ionisation front over the course of
our simulation which we retrieved from snapshots taken at regular time
intervals. The position of the IF is determined by calculating the mean radius
where the shell-averaged ionised medium drops to 50 \%, signalling the rapid
transition to the neutral medium. The result is shown in
\Figref{fig:starbench-if-over-time} together with the Starbench
solution \eqref{eq:starbench} (black solid line) and the Spitzer solution
\eqref{eq:spitzer} (grey dashed line). Clearly, our numerical solution matches
the Spitzer solution only at very early times, but there is good agreement to
the Starbench solution throughout the whole simulation. We ascribe the
deviation observable from $t = 1.1\,\text{Myr}$ on-wards to the bubble nearing
the boundaries of the domain and the still rather low resolution.
Investigations with identical parameters but different solvers in FLASH, such
as PPM, revealed the exact same behaviour, hence the phenomenon is not rooted
in our fluid solver.

From a technical point of view, this particular test setup is remarkable, since
it unites a large number of numerical components into one
simulation. The complex interplay of physics and chemistry, especially the
correct handling of a very sharp transition of a two-component
(ionised/neutral) medium intertwined with ionisation, recombination, heating
and cooling is a major challenge for every fluid dynamics code. 

In our next section, we extend the Starbench test setup with an
initial fractal density distribution to get a more realistic picture of a
typical D-type expansion seen in astronomical observations.

\subsubsection{D-type Expansion into a Fractal Molecular Cloud}
\label{sec:fractal}
Radiation driven feedback from massive stars is believed to be a key element in
the evolution of molecular clouds. In this work, we utilise our new DGFV4 solver
to model the dynamical effects of a single O7 star emitting ionising photons
at $10^{49}\,\text{s}^{-1}$ located at the centre of a spherical molecular cloud. The
cloud consists of $10^4\,\text{M}_{\odot}$ and stretches over a region of $6.4
\,\text{pc}$ in each direction, so the volume-weighted mean density of the
cloud amounts to $\rho_0 = 0.62 \cdot 10^{-21}\,\text{g}\,\text{cm}^{-3}$.

It is a well established fact that molecular clouds are rich in internal
structure probably driven by pure turbulence
\citep{klessen2011star,girichidis2011importance} and
subscribe to a statistically self-similar fractal structure
\citep{stutzki1998fractal}.  Our goal is to model the development of a molecular
cloud having initial fractal dimension of $\mathcal{D} = 2.6$ with a standard
deviation of the approximately log-normal PDF of $\sigma_{\text{STD}} = 0.38$.
With these parameters the cloud is initially dominated by small-scale
structures, which are then quickly overrun by the advancing ionisation front,
thereby producing neutral pillars protruding into the HII region. The ionised
gas within the expanding hot bubble blows out through a large number of small
holes between these pillars. These regions are termed pillar-dominated
\citep{walch2012dispersal}.

Fractals with mass-size relations similar to those observed by
\citet{larson1981turbulence} can be created in Fourier space
\citep{shadmehri2011mass}, so we can freely configure the fractal dimension
$\mathcal{D}$ of the initial cloud. A detailed account on how to construct such
a fractal density distribution is presented in \citet{walch2012dispersal}.

\begin{figure}
\centering
\includegraphics[width=1.0\columnwidth]{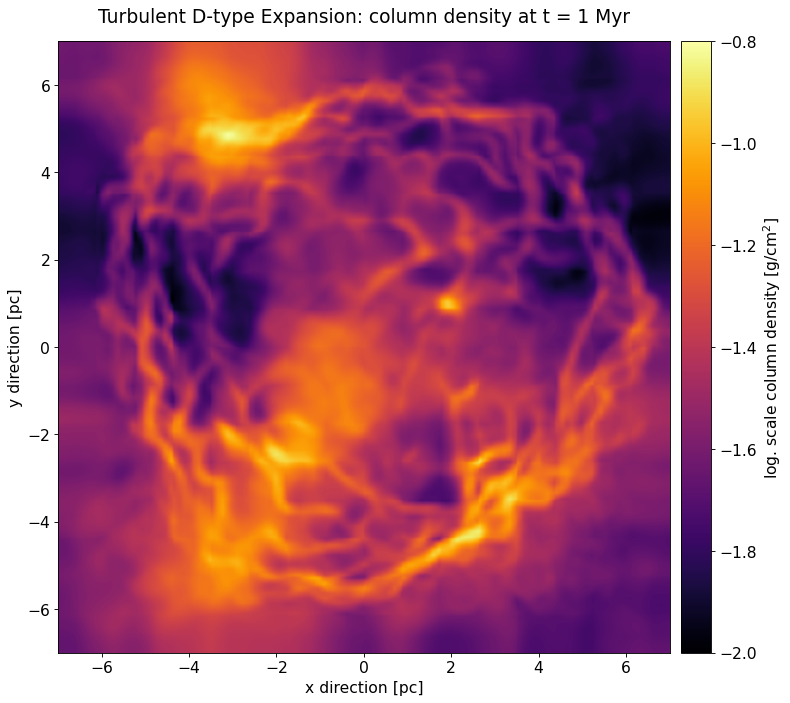}
\caption{Turbulent D-type expansion into a fractal medium run by our DGFV4
solver. Shown is the column density along the z-axis: 
$\int_{z_{\mathrm{min}}}^{z_{\mathrm{max}}}\,\rho(x,y,z)\,\mathrm{d}z$.}
\label{fig:turb-expansion-column-density}
\end{figure}

The aspects of radiation physics are handled by the TreeRay module in FLASH
which we introduced in detail before. Besides minor changes in setup parameters
given below, this setup uses the same parameters as in \Secref{sec:starbench}.
In contrast to \Secref{sec:starbench}, we simulate the whole expanding bubble,
but keep the same maximum resolution at $128^3$ cells. Hence, the radiation
source is at the centre of the cubic computational domain $\Omega =
[-7\,\text{pc},7\,\text{pc}]^3$ and all boundaries are set to outflow. 
Radiative cooling effects of the molecular cloud itself are neglected.
The result at final simulation time $T = 1\,\text{Myr}$ can be seen in
\Figref{fig:turb-expansion-column-density}.  The figure shows the column
density in z-direction uncovering the aforementioned pillars as a product of
the turbulent density distribution tearing apart the shell structure.

Since the setup combines multi-species and radiation physics in a very
tough shock-turbulence regime, this final simulation shows that our new DG
implementation in FLASH is capable of running complex multi-physics applications
in astrophysical settings.

\section{Conclusions}
In this paper, we discuss the implementation of a high-order DG fluid dynamics
solver within the simulation framework FLASH, with a specific focus on
multi-physics problems in astrophysics. We focus in particular on the
modifications of the DG scheme to make it fit seamlessly within FLASH in such a
way, that all other physics module can be naturally coupled without additional
implementation overhead. A key to this is that our DG scheme uses mean value
data organised into blocks. In specialised sections on different parts of the
solver, we explain the strategy and compromises that we use to get a feasible
implementation of the DG solver with the existing module interfaces of FLASH.
The compromises are mainly related to accepting that the other physics modules
assume mean value data (and not polynomial data), which means that some parts
of the coupling are limited to second order accuracy, e.g., the evaluation of
the gravitational source terms. 

We demonstrate in extensive numerical tests that the novel DG implementation in
FLASH works properly and is fully connected to all multi-physics modules. Step
by step, the complexity of the test cases is increased by using setups that
require more complex physical modules and features. Our novel DG solver in FLASH
is ready for use in astrophysical simulations and thus ready for assessments and investigations.

\section*{Acknowledgements}

JM, GG and SW acknowledge funding through the Klaus-Tschira Stiftung via the
project "DG$^2$RAV". JM and GG thank the European Research Council for funding
through the ERC Starting Grant “An Exascale aware and Un-crashable
Space-Time-Adaptive Discontinuous Spectral Element Solver for Non-Linear
Conservation Laws” (EXTREME, project no. 71448). JM thanks the FLASH Center for Computational Science
at the University of Chicago for providing a copy of the FLASH code.
SW thanks the DFG for funding through SFB~956 ``The conditions and impact of star formation'' (sub-project C5) and gratefully acknowledges funding from the European Research Council via
the ERC Starting Grant "The radiative interstellar medium" (project number
679852) under the European Community's Framework Programme FP8. This work was
performed on the Cologne High Efficiency Operating Platform for Sciences
(CHEOPS) at the Regionales Rechenzentrum K\"oln (RRZK) and on the group cluster ODIN.
We thank RRZK for the hosting and maintenance of the clusters.

%%%%%%%%%%%%%%%%%%%%%%%%%%%%%%%%%%%%%%%%%%%%%%%%%%
\section*{Data Availability}
The code is open source and can be accessed under \href{https://github.com/jmark/DG-for-FLASH}{github.com/jmark/DG-for-FLASH}.



\bibliographystyle{mnras}
\bibliography{citations} % if your bibtex file is called example.bib

%\begin{thebibliography}{99}
%\end{thebibliography}

%%%%%%%%%%%%%%%%%%%%%%%%%%%%%%%%%%%%%%%%%%%%%%%%%%

%%%%%%%%%%%%%%%%% APPENDICES %%%%%%%%%%%%%%%%%%%%%
\appendix

\section{Comparative Study for the MHD Torus Setup}
\label{sec:comp-study-torus}
Here we present a small comparative study of the MHD torus setup
introduced in \Secref{sec:mhd-torus}. We run a total of
six simulations: three different resolutions of $64^3$, $128^3$, and $256^3$ cells
each with the two MHD solvers Bouchut5 \citep{bouchut2010multiwave} and DGFV4.
We use Bouchut5, a well-tested, second order, split solver in FLASH,
since the standard solver for MHD, an unsplit, staggered mesh solver \citep{lee2013solution}, crashes for this setup.
As time integrator in DGFV4 we choose the third order, four stages SSP-RK(4,3) \citep{spiteri2002new} scheme. All runs
are conducted with the ODIN cluster (hosted at the computing centre at the University of Cologne) on 16 nodes each equipped with 16
Intel Xeon CPUs model E5-2670 (2.6 GHz). This gives a total cpu count (\#cpus) of 256 cores.

\begin{figure*}
\centering
\includegraphics[width=0.9\textwidth]{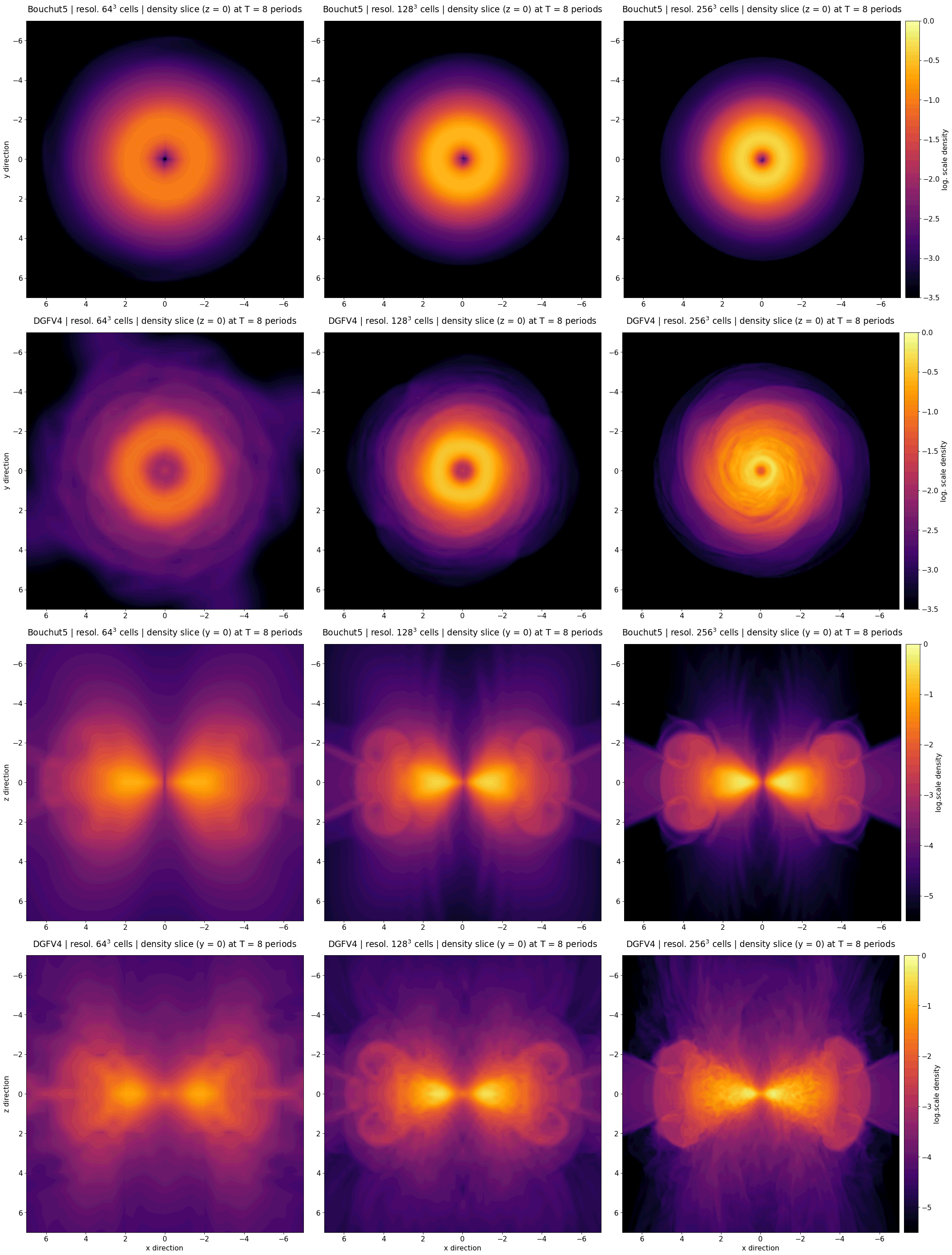}
\caption{MHD torus after 8 periods run for max. resolutions (from left to right)
$64^3$, $128^3$, $256^3$ and two fluid solvers Bouchut5 (rows 1,3) and DGFV4
(rows 2,4). Shown are log-scale density slices in the x-y plane (rows 1,2)
and in the x-z plane (rows 3,4).}
\label{fig:torus-dens}
\end{figure*}

\begin{figure*}
\centering
\includegraphics[width=0.9\textwidth]{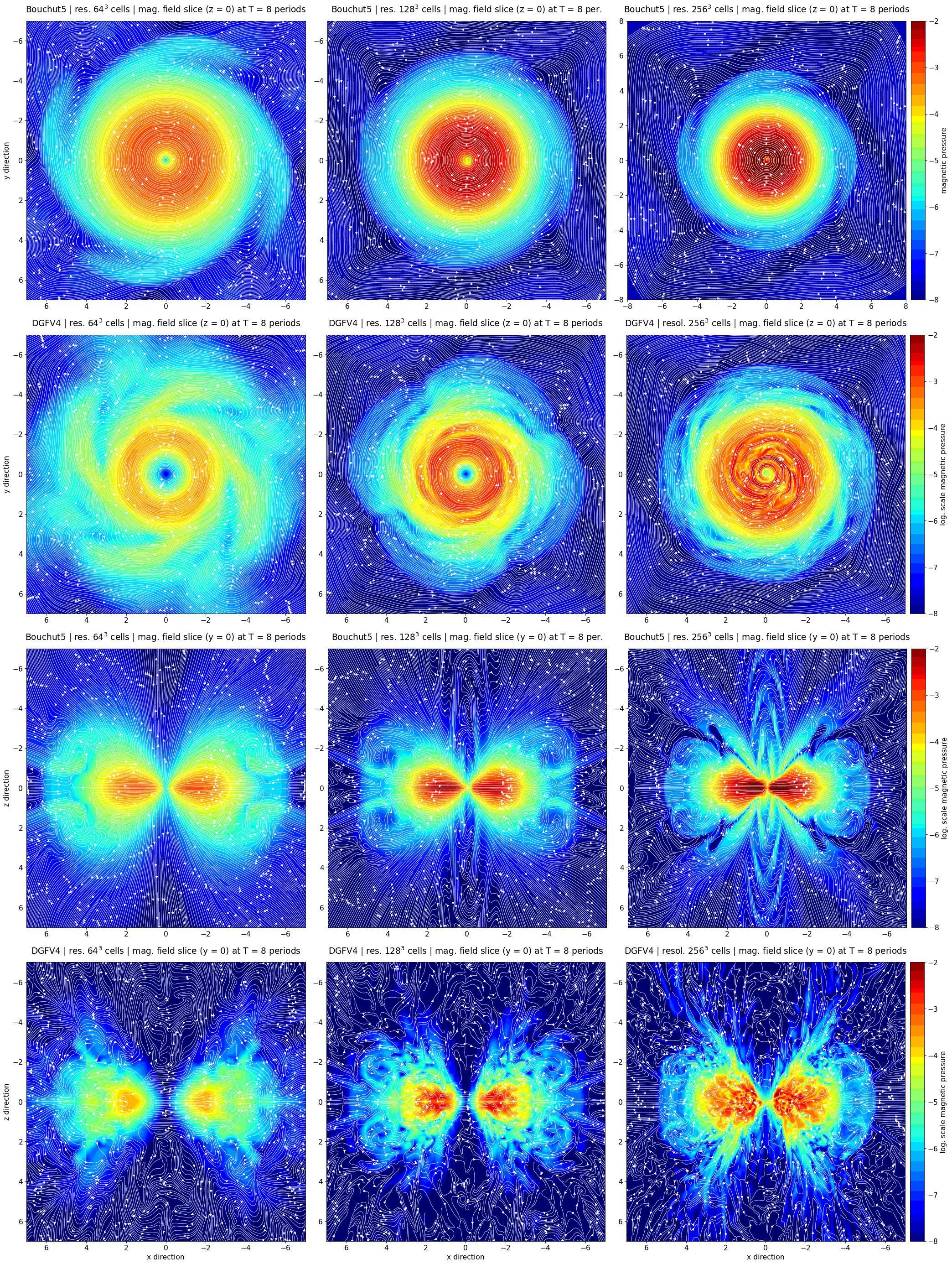}
\caption{MHD torus after 8 periods run for max. resolutions (from left to right)
$64^3$, $128^3$, $256^3$ and two fluid solvers Bouchut5 (rows 1,3) and DGFV4
(rows 2,4). Shown are log-scale magnetic pressure slices in the x-y plane (rows 1,2)
and in the x-z plane (rows 3,4). The streamlines in white denote the
magnetic field lines.}
\label{fig:torus-magp}
\end{figure*}

The results at final simulation time are shown in \Figref{fig:torus-dens}
and \Figref{fig:torus-magp}. Clearly, the DGFV4 solver shows
finer structures and more scales in the disk at every resolution level compared to Bouchut5
and is capable to resolve magneto-rotationally driven outflows at lowest resolution levels.

In Table~\ref{tab:mhd-torus-runtimes} we list the runtimes of all six simulations. Bouchut5 is a split solver and thus in general has a larger time step (about a factor of 2) compared to unsplit schemes. Hence, the number of time steps  (\#steps) are lower for Bouchut5 compared to DGFV4, which results in overall faster completion times for the same resolution. To further assess our implementation, we compute a performance index (PID) as
\begin{equation}
\label{eq:PID}
    \mathrm{PID} = (\mathrm{runtime}\,\times\,\#\mathrm{cpus})\,/\, (\#\mathrm{cells}\,\times\,\#\mathrm{steps})
\end{equation}
which we also list in Table~\ref{tab:mhd-torus-runtimes}. PID is basically the average cost of a spatial cell per time step. A lower PID means better performance. Note, that the MPI-parallelised code scales better with increasing resolution resulting in lower PIDs for both schemes. A drawback of DG schemes compared to FV discretizations is, that the spatial operator spectra are more stiff and hence need more stable time integration methods, such as the high-order RK method used in this comparison. This means, that the DGFV4 discretization has to be called four times per time step. Factoring this in and comparing the CPU time per evaluation of the spatial operator gives an unexpected result. The novel DGFV4 implementation is quicker than the Bouchut5 implementation, as one has to divide the PID of DGFV4 by a factor of four to get the cost of a spatial grid cell per 'sweep'.

\begin{table}[h!]
\centering
\caption{Runtimes of the MHD torus setup introduced in \Secref{sec:mhd-torus} for the Bouchut5 and our DGFV4 solver
implemented in FLASH (v4.3.). All simulations run on the cluster ODIN
on 16 nodes á 16 cores. The definition for the performance index (PID)
is given by eqn. \eqref{eq:PID}.}
\label{tab:mhd-torus-runtimes}
\begin{tabular}{crrrr}
\hline
solver & \#cells & runtime [h] & \#steps & PID [$\mu$s] \\
 \hline
 \multirow{3}{*}{Bouchut5} &
   $64^3$  &      0.03 &      940 &        110.00 \\
 & $128^3$ &      0.41 &     3012 &        60.16 \\
 & $256^3$ &      6.14 &     8374 &        40.26 \\
\hline
\multirow{3}{*}{DGFV4} &
   $64^3$  &   0.08 &     1736 & 156.10 \\
 & $128^3$ &   1.18 &     4808 & 107.80 \\
 & $256^3$ &  17.24 &    13898 & 68.16 \\
\hline
\end{tabular}
\end{table}

%%%%%%%%%%%%%%%%%%%%%%%%%%%%%%%%%%%%%%%%%%%%%%%%%%

% Don't change these lines
\bsp	% typesetting comment
\label{lastpage}
\end{document}